\documentclass[12pt]{article}
\pdfoutput=1

\usepackage{amsmath,amsfonts,amssymb,epsfig}
\usepackage{slashed}
\usepackage{cite}
\usepackage[width=0.9\textwidth,margin={30pt,30pt},font=small,labelfont=bf]{caption}
\numberwithin{equation}{section}

\usepackage{hyperref}

\usepackage{xcolor}
\hypersetup{
    colorlinks,
    linkcolor={red!0!black},
    citecolor={blue!50!black},
    urlcolor={blue!80!black}
}

\usepackage{graphicx}
\usepackage{cancel}
\usepackage{cite}
\usepackage{color}
\usepackage{multirow}

\usepackage{xspace}

\newcommand{\SARAH}{{\tt SARAH}\xspace}

\newcommand{\scr}[1]{\ensuremath{\mathcal{#1}}}
\newcommand{\exclude}[1]{}
\def\nn{\nonumber}

\def\bra{\langle}
\def\ket{\rangle}

\def\beq{\begin{equation}}
\def\eeq{\end{equation}}
\def\bal{\begin{align}}
\def\eal{\end{align}}

\def\s2b{s_{2\beta}}
\def\c2b{c_{2\beta}}

\def\gtu{\tilde{g}_{u}}
\def\gtd{\tilde{g}_{d}}
\def\gtup{\tilde{g}'_{u}}
\def\gtdp{\tilde{g}'_{d}}

\def\L{\mathcal{L}}

\def\micrOMEGAs{{\tt micrOMEGAs}}
\def\SARAH{{\tt SARAH}}

\def\2b2[#1,#2][#3,#4]{\left( \begin{array}{cc} #1 & #2 \\ #3 & #4 \end{array}
\right)}
\def\3b3[#1,#2,#3][#4,#5,#6][#7,#8,#9]{\left( \begin{array}{ccc} #1 & #2 &#3 \\
#4 & #5 & #6\\#7&#8&#9\end{array} \right)}
\def\thv[#1,#2,#3]{\left( \begin{array}{c} #1 \\ #2 \\ #3 \end{array} \right)}
\def\twv[#1,#2]{\left( \begin{array}{c} #1 \\ #2 \end{array} \right)}

\def\ov{\overline}

\def\smallW{{\scriptscriptstyle W}}
\def\smallB{{\scriptscriptstyle B}}

\def\MS{M_S}

\def\tBp{{\tilde \smallB^\prime}}
\def\tWp{{\tilde \smallW^\prime}}

\def\mtBp{m_\tBp}
\def\mtWp{m_\tWp}

\def\msbar{{\ov {\rm MS}}}
\def\beq{\begin{equation}}
\def\eeq{\end{equation}}
\def\bea{\begin{eqnarray}}
\def\eea{\end{eqnarray}}

\def\matw[#1,#2][#3,#4]{\left(\begin{array}{cc} #1 & #2 \\ #3 & #4 \end{array}\right)}
\def\matws[#1,#2,#3]{\left(\begin{array}{cc} #1 & #2 \\ #2 & #3 \end{array}\right)}

\def\nn{\nonumber}

\def\reps[#1,#2,#3]{$(\mathbf{#1},\mathbf{#2},#3)$}
\def\hE{\hat{E}}
\def\hEt{\hat{\tilde{E}}}

\def\bra{\langle}
\def\ket{\rangle}

\def\beq{\begin{equation}}
\def\eeq{\end{equation}}
\def\bal{\begin{align}}
\def\eal{\end{align}}

\def\s2b{s_{2\beta}}
\def\c2b{c_{2\beta}}

\def\L{\mathcal{L}}

\def\2b2[#1,#2][#3,#4]{\left( \begin{array}{cc} #1 & #2 \\ #3 & #4 \end{array}
\right)}
\def\3b3[#1,#2,#3][#4,#5,#6][#7,#8,#9]{\left( \begin{array}{ccc} #1 & #2 &#3 \\
#4 & #5 & #6\\#7&#8&#9\end{array} \right)}
\def\thv[#1,#2,#3]{\left( \begin{array}{c} #1 \\ #2 \\ #3 \end{array} \right)}
\def\twv[#1,#2]{\left( \begin{array}{c} #1 \\ #2 \end{array} \right)}

\def\ov{\overline}

\def\WDM{ \tilde{W} |_{\text{DM}}}
\def\BHDM{ \tilde{B}/\tilde{H} |_{\text{DM}}}
\def\HDM{ \tilde{H} |_{\text{DM}}}

\def\Eunif {{E^\prime}}

\def\ehat{\varepsilon}

\newenvironment{Appendix}
 {
  \setcounter{section}{0}
  \setcounter{equation}{0}

 }

\begin{document}

\begin{flushright}
\end{flushright}
\begin{center}

\vspace{1cm}
{\LARGE{\bf (O)Mega Split
 }}

\vspace{1cm}

\large{  Karim Benakli$^\clubsuit$ \let\thefootnote\relax\footnote{$^\clubsuit$kbenakli@lpthe.jussieu.fr},
Luc Darm\' e$^\heartsuit$ \footnote{$^\heartsuit$darme@lpthe.jussieu.fr},
Mark~D.~Goodsell$^\diamondsuit$ \footnote{$^\diamondsuit$goodsell@lpthe.jussieu.fr}
 \\[5mm]}

{\small
\emph{1-- Sorbonne Universit\'es, UPMC Univ Paris 06, UMR 7589, LPTHE, F-75005, Paris, France \\
2-- CNRS, UMR 7589, LPTHE, F-75005, Paris, France \\
}}

\end{center}
\vspace{0.7cm}

\abstract{We study two realisations of the Fake Split Supersymmetry Model (FSSM), the simplest model that can easily reproduce the experimental value of the Higgs mass for an arbitrarily high supersymmetry scale $\MS$, as a consequence of swapping  higgsinos for equivalent states, fake higgsinos, with suppressed Yukawa couplings. If the LSP is identified as the main Dark matter component, then a standard thermal history of the Universe implies upper bounds on $\MS$, which we derive. On the other hand, we show that renormalisation group running of soft masses \emph{above} $\MS$ barely constrains the model -- in stark contrast to Split Supersymmetry -- and hence we can have a ``Mega Split'' spectrum even with all of these assumptions and constraints, which include the requirements of a correct relic abundance, a gluino life-time compatible with Big Bang Nucleosynthesis and absence of signals in present direct detection experiments of inelastic dark matter. In an appendix we describe a related scenario, Fake Split Extended Supersymmetry, which enjoys similar properties.}

\newpage

\tableofcontents

\setcounter{footnote}{0}

\section{Introduction}
\label{SEC:INTRO}

There is good reason to believe that supersymmetry plays a fundamental role in nature at some
energy scale, but there is increasing concern that it may not fully solve the hierarchy problem.
However, one key hint to its relevance is the apparent unification of gauge couplings, and if we 
take this as the main \emph{phenomenological} motivation, accepting fine-tuning of the 
electroweak scale -- since the fine-tuning of one mass in the Higgs potential has perhaps an anthropic justification
-- then we are led to study theories where unification
arises naturally without tuning of other particle mass thresholds, since the apparent unification of couplings 
could have no other explanation. This reasoning led to much study of Split Supersymmetry (Split SUSY) \cite{ArkaniHamed:2004fb, GiudiceRomanino, ArkaniHamed:2004yi}; of particular relevance to this work from the burgeoning literature are \cite{arvanitaki_limits_2005,minisplit,bernal_mssm_2007,Giudice:2011cg,Bagnaschi:2014rsa,Vega:2015fna}. 

In \cite{GiudiceRomanino}
the set of conditions for generic theories extending the Standard Model that predict gauge coupling unification naturally were
considered, and the simplest among these theories where the only new particles near the electroweak scale are fermions was argued to be
Split SUSY. The requirement of no new light scalars might seem at first to be ad-hoc, but without requiring an unjustifiable fine-tuning
it is difficult to include such fields, leading to much more complicated theories -- whereas fermion
masses can be easily protected by approximate continuous symmetries.

However, the conclusion of minimality for Split SUSY only applies to the spectrum of particles. 
Although we would like to impose the requirement that the theory above some high energy scale $\MS$ is supersymmetric --
since we insist on a fundamental role for SUSY in nature -- in the literature it has also 
been assumed, implicitly or explicitly, that the high-energy theory is the MSSM. This has a number of consequences
and drawbacks as we shall review below. Here, extending previous work \cite{dudas_flavour_2013,FSSM1}, we shall consider
different high-energy theories which yield the same low-energy particle content but with different couplings
which allow the drawbacks of Split SUSY to be alleviated. 

The first drawback concerns the observed value of the Higgs mass: with the MSSM as the 
high-energy theory, in Split SUSY the maximum value for $\MS$ allowed to be compatible with this one
constraint is $10^8$ GeV -- worse than high-scale SUSY (where the theory below $\MS$ is the Standard Model) which allows 
$\MS$ up to $10^{12}$ GeV. This constraint arises from the point at which the Higgs quartic coupling runs negative. 
This is a problem if one would like to avoid introducing additional intermediate scales, and have soft masses at the unification
or Planck scale; or alternatively to explain a common scale with the QCD axion or even right-handed neutrinos. However, a more severe problem arises
when we consider the effect of running of the soft masses above $\MS$; as explained in \cite{minisplit} and as we shall briefly review
in section \ref{SEC:RESULTS}, without additional unjustifiable tuning the tangent of the mixing angle of the MSSM Higgs bosons at $\MS$ should be rather different
from $1$, and as a consequence the maximum value for $\MS$ consistent with the obvervable Higgs mass should be considered to be 
much lower, around $10^5$ or $10^6$ GeV -- a ``Mini-Split''~\cite{minisplit}. The final drawback is that unification of gauge couplings provides in general a constraint that the soft masses should not be 
generated at a scale too far above $\MS$, putting -- for example -- gravity-mediation scenarios into tension. 

In previous work \cite{FSSM1} it was shown that by changing the theory above $\MS$ the relationship between the Higgs mass 
and the scale $\MS$ can be completely changed. One particular theory was defined, which below $\MS$ was named the ``Fake Split Supersymmetric Model'' (FSSM) 
since it has the same particle content as Split SUSY but where the non-Standard Model Yukawa couplings involving the new fermions are suppressed. In this model it was shown that, in fact, the observed value of the Higgs mass arises very naturally for \emph{any} value of $\MS$. The scenario arose naturally by simply changing the theory above $\MS$ to a unified model inspired by Dirac gaugino phenomenology. 

Encouraged by this success, in this work, after reviewing the low-energy theory and our original scenario in section \ref{SEC:FSSM},
we shall provide a second realisation of the FSSM with subtly different and improved phenomenological properties, that derives from a much 
simpler extension of the MSSM above $\MS$; it requires simply two vector-like pairs of $SU(5)$ fundamentals/antifundamentals. Both theories
enjoy the same prediction for the Higgs mass, but in section \ref{SEC:RESULTS} we shall examine how both scenarios fare when we include possible constraints from running the soft masses above $\MS$. We shall show that this \emph{barely constrains the scenario at all}. Furthermore, to add the icing on the cake
we shall consider the constraints from assuming a standard cosmology and the consequent predictions for dark matter, showing that even under this tight straightjacket the FSSM can be consistent with a high supersymmetry scale of $10^8$ to $10^{10}$ GeV -- a ``Mega Split,'' potentially related to the QCD axion scale -- and  completely consistent with mediation at \emph{any} scale above $\MS$. 

Finally, in appendix \ref{APP:FSES} we describe a twist on the Fake Split SUSY scenario, \emph{Fake Split Extended Supersymmetry}, which enjoys similar properties to the FSSM -- but may also have some connection with the recently discussed X-ray line at $3.55$ keV. It is relatively self-contained and so readers interested only in the line can read that independently.

\section{Fake Split Supersymmetry Models}
\label{SEC:FSSM}

 In this section we describe two realisations of the FSSM. In the FSSM-I, both fake higgsinos and gauginos are introduced, as in~\cite{FSSM1} and~\cite{dudas_flavour_2013}, while in the FSSM-II, only the higgsino-like light fermions are fake.

However, the particle content of the FSSM below $\MS$ is the same as in Split SUSY; it contains SM fields plus  a set of fermions with quantum numbers of the higgsinos and gauginos. It differs in the fact that the Yukawa couplings of these non-SM fermions with the light Higgs boson do not obey the same constraint at $\MS$. We are interested here in the case where these couplings are suppressed, which, as shown in \cite{FSSM1}, can be consistent with the observed Higgs mass for any value of $\MS$. 

The fake higgsino-like particles (F-higgsinos) $\tilde{H}_{u,d}^\prime $ and gaugino-like  particles (gauginos or F-gauginos) $\tilde{B}^\prime$, $\tilde{W}^\prime$ and $\tilde{g}^\prime$ have couplings with the Higgs in the low-energy Lagrangian of
\begin{align}
\label{eq:gtildas}
  & {\cal L}_{\rm eff} \supset -\frac{H^{\dagger}}{\sqrt{2}}
  (\tilde{g}_{2u}\, \sigma^a\, \tilde {W^\prime}^a + \tilde{g}_{1u}\, \tilde
  {B}^\prime) \ \tilde {H}_u^\prime - \frac{H^{T} i \sigma^2}{\sqrt{2}}
  (- \tilde {g}_{2d} \,\sigma^a\, \tilde {W^\prime}^a + \tilde{g}_{1d}\,
  \tilde{B}^\prime) \ \tilde{H}_d^\prime~.
\end{align}
In the models considered below the coupling constants $\tilde{g}_{1u} , \tilde{g}_{1d},\tilde{g}_{2u}, \tilde{g}_{2d}$ are  suppressed by a power of a small parameter $\varepsilon$, arising from the breaking of an approximate symmetry. We shall consider two realisations of the FSSM in the following, with different origins of (and parametric dependence on)  $\varepsilon$: from an additional approximate $U(1)_F$ ``flavour'' symmetry in the FSSM-I,  and  from an approximate R-symmetry in FSSM-II. These two models will be described in detail in the next subsections. 

However, both versions of the FSSM make the same prediction for the Higgs quartic coupling at $\MS$ \emph{at tree level} as split SUSY:
\beq
\label{lambdaMS}
\lambda (\MS)~=~
\frac14\left(g^2+g^{\prime\,2}\right)\,\cos^22\beta ~+~ \Delta^{(\ell)} \lambda ~+~ \Delta^{(\ov{MS})} \lambda~+~{\cal
  O}(\varepsilon^2)~. 
\eeq
In this work the subleading corrections in $\varepsilon$ will always be negligible. More important are the loop contributions $\Delta^{(\ell)} \lambda$ (and less so the conversion between $\ov{MS}$ and $\ov{DR}$ written as $\Delta^{(\ov{MS})} \lambda$); these differ between the FSSM-I and FSSM-II, and more discussion about the estimation of their contributions can be found in Appendix~\ref{APP:implementation}.

\subsection{Type I FSSM}

The original FSSM construction (for short FSSM-I) arose from the framework of Dirac Gauginos  \cite{FSSM1, dudas_flavour_2013} where a chiral superfield in the adjoint representation is added for each gauge group. The field content of the FSSM-I in the UV is actually similar to the MDGSSM of~\cite{benakli_constrained_2014} albeit with a different mass hierarchy as we will see below. We review here the main points of this construction. In the following bold-face symbols denote superfields. 

The adjoint chiral superfields are called ``fake gauginos'' (henceforth F-gauginos). They consist of a set of chiral multiplets, namely a singlet $\mathbf{S} = S +
  \sqrt{2} \theta \chi_{S }+ \ldots$ ; an $SU(2)$ triplet $\mathbf{T}
  = \sum_a \mathbf{T}^{a} \,\sigma^a/2$, where $ \mathbf{T}^{a} =
  T^{a} + \sqrt{2} \theta \chi_T^{a} + \ldots $ where $\sigma^a$ are the
  three Pauli matrices; and an $SU(3)$ octet $\mathbf{O} = \sum_a
  \mathbf{O}^a\, \lambda^a/2$, where $\mathbf{O}^a = O^{a} + \sqrt{2}
  \theta \chi_O^{a}+ \ldots$ and $\lambda^a$ are the eight Gell-Mann
  matrices.

Unification is jeopardised if one does not add further fields since the F-gaugino multiplets do not fill complete representations of a GUT group. An easy way to recover unification is to add two pairs of vector-like right-handed electron superfields ($\mathbf{{E'}}_{1,2}$ in $(\mathbf{1} ,\mathbf{1})_{1}$ and $\mathbf{\tilde{{E'}}}_{1,2}$ in $  (\mathbf{1} ,\mathbf{1})_{-1}$) and one pair of  $SU(2)$ doublets ($\mathbf{H_d^\prime}$ in $(\mathbf{1} ,\mathbf{2})_{1/2}$ and $\mathbf{H_u^\prime}$ in $(\mathbf{1} ,\mathbf{2})_{-1/2}$).
In this work, the latter become fake Higgs doublets (henceforth F-Higgs) and their fermionic components fake higgsinos (henceforth F-higgsinos) rather than, for example, assigning them lepton number (as in \cite{benakli_constrained_2014}). 

An essential difference with usual Dirac gaugino models is that we do not impose an $R$-symmetry which forbids Majorana gaugino masses leading to the same mass for gauginos and F-gauginos. Instead, we keep only the F-gauginos light thanks to an approximate $U(1)_F$ flavour symmetry with the following charge assignments
\begin{center}
\begin{tabular}{|c|c|}\hline
Superfield & $U(1)_F$ charge \\\hline
$\mathbf{H_u^\prime}, \mathbf{H_d^\prime}$; $\mathbf{S}, \mathbf{T}, \mathbf{O}$& $1$ \\ \hline
$\mathbf{{\Eunif}}_{1,2}, \mathbf{\tilde{{\Eunif}}}_{1,2}$ & $0$ \\ \hline
\end{tabular}\end{center}
All other (MSSM) multiplets are neutral under $U(1)_F$. We parametrise the breaking of this symmetry by a small number $\varepsilon$ which could be considered, as standard in flavour models, to come from the expectation value of a field (divided by some UV scale); in this case we can suppose it to have charge $-1$ under $U(1)_F$.

The superpotential contains a hierarchy of couplings due to suppressions by different powers of $\varepsilon$:
\begin{eqnarray}
\label{SuperpotentialFSSM}
W &\supset& W_{\text{unif}} +  \mu_0 \, \mathbf{H_u\cdot H_d } 
+ Y_u \, \mathbf{U^c \, Q \cdot H_u} 
- Y_d \, \mathbf{D^c \, Q \cdot H_d} 
- Y_e \, \mathbf{E^c \, L \cdot H_d} \nn\\
& +& \varepsilon \,\left(
\hat{\mu}'_d \,\mathbf{H_u\cdot H_d^\prime } + 
\hat{\mu}'_u \,\mathbf{H_u^\prime \cdot H_d }
+ \hat{Y}_u^\prime\, \mathbf{U^c\,Q \cdot H_u^\prime} 
- \hat{Y}_d^\prime\, \mathbf{D^c\,Q \cdot H_d^\prime} 
- \hat{Y}_e^\prime\, \mathbf{E^c\,L \cdot H_d^\prime}\right)
\nn\\
& +&  \varepsilon\, \left(
\hat{\lambda}_S \, \mathbf{S \, H_u\cdot H_d}  
+ 2\, \hat{\lambda}_T\, \mathbf{H_d\!\cdot\! T\, H_u}\right) 
\nn\\
& +& \varepsilon^2 \,\left(
\hat{\lambda}_{Sd}^\prime\, \mathbf{S \,H_u\cdot H_d^\prime }  
+ \hat{\lambda}_{Su}'\, \mathbf{S \,H_u^\prime\cdot H_d } 
+ 2\, \hat{\lambda}_{Tu}^\prime \, \mathbf{H_d\!\cdot\! T \,H_u^\prime} 
+ 2\, \hat{\lambda}_{Td}^\prime\, \mathbf{H_d^\prime\!\cdot\! T \,H_u}
\right)
\nn\\
& +& 
\varepsilon^2\,\hat{\mu}''\, \mathbf{H_u^\prime \cdot H_d^\prime }
~+~
\varepsilon^2 \,\left[
\frac12 \,\hat{M}_S \,\mathbf{S}^2 + \hat{M}_T\,
\textrm{Tr}(\mathbf{TT}) + \hat{M}_O \,\textrm{Tr}(\mathbf{OO}) \right] + \scr{O}(\varepsilon^3)~,
\end{eqnarray}
where $\mathbf{Q, U^c , D^c, L \text{ and }E^c}$ are the quarks and leptons superfields, $\mathbf{H_u \text{ and } H_d}$ the usual MSSM two Higgs doublets. We have explicitly written the $\varepsilon$ factors so that all mass parameters are expected to be generated at $\MS$ and all dimensionless couplings are either of order one or suppressed by loop factors. 
The additional superpotential $W_{\text{unif}}$ contains the interactions involving the pairs $\mathbf{\Eunif}_{1,2}$ and $\mathbf{\tilde{\Eunif}}_{1,2}$; these fields are irrelevant for the low energy theory because their masses are not protected, so are of order $\MS$. 

We shall not explicitly write all of the soft terms in the model for reasons of brevity, since they can simply be inferred from the flavour assignments.
For example, for the gauginos, allowing all terms permitted by the symmetries we have unsuppressed Majorana masses for the gauginos, and then the suppressed Majorana masses for the F-gauginos $\varepsilon^2 \hat{M}_{S,T,O}$ -- and Dirac masses mixing the two suppressed only by $\varepsilon$, giving a generic mass matrix of  
\beq
\label{eq:ginomass1}
\mathcal{M}_{1/2}~\sim~ \scr{O}(\MS) \, \begin{pmatrix}
1    &   \mathcal{O}(\varepsilon)  \\
  \mathcal{O}(\varepsilon) &\mathcal{O}(\varepsilon^2)
\end{pmatrix}.
\eeq
We have a heavy eigenstate of mass ${\cal O}(\MS)$ and a light one,  the F-gaugino at leading order, of
mass ${\cal O}(\varepsilon^2 \MS)$. 
Requiring that the F-gauginos have a mass at the TeV scale (for unification and, as we shall later see, dark matter) then fixes $\varepsilon$:
\begin{align}
 \varepsilon = \scr{O}(\sqrt{\frac{ \text{TeV}}{\MS}}) \ .
\end{align}
For the adjoint scalars we shall define the explicit soft terms:
\begin{align}
- \L_{\text{scalar soft}}  \supset &\  m_S^2 |S|^2  + 2 m_T^2 \text{tr} \, T^\dagger T   + 2 m_O^2 \text{tr} \, O^\dagger O  \nn\\
&+ \frac{1}{2} \varepsilon^2 B_S [ S^2 + h.c] +   \varepsilon^2 B_T [ \text{tr} \, T T + h.c] +  \varepsilon^2 B_O  [ \text{tr} \, OO  + h.c] \ . 
\end{align} 
We see that the $B$ parameters are $\varepsilon^2$-suppressed, circumventing a common feature in some Dirac Gaugino models of predicting tachyonic scalar adjoints.

The Higgs mass matrix can be written in terms of the four-vector\\$ v_H\equiv (H_u, {H_d}^*,
H_u^\prime, H_d^{\prime\,*})$ as
\begin{align}
-\frac{1}{\MS^2}\mathcal{L}_{soft} ~~\supset ~~& v_H^\dagger \left(
\begin{array}{cccc} \mathcal{O}(1) & \mathcal{O}(1) & \mathcal{O}(\varepsilon) &
\mathcal{O}(\varepsilon) \\ \mathcal{O}(1) &
\mathcal{O}(1) & \mathcal{O}(\varepsilon) &
\mathcal{O}(\varepsilon) \\ \mathcal{O}(\varepsilon) & \mathcal{O}(\varepsilon)
& \mathcal{O}(1) & \mathcal{O}(\varepsilon^2) \\
\mathcal{O}(\varepsilon) &
\mathcal{O}(\varepsilon)& \mathcal{O}(\varepsilon^2) & \mathcal{O}(1)
\end{array} \right) v_H .
\end{align}
In the spirit of Split SUSY we tune the weak scale to its correct value and define the SM-like Higgs boson $H$ as 
\beq
\label{eq:lincomb1}
H_u ~\approx~ \, \sin\beta\, H\, + \ldots~, 
~~~ \qquad H_d ~\approx~ \,
\cos\beta\,i\sigma^2\, H^* \,+ \dots~,
\eeq
\beq
\label{eq:lincomb2}
H_u^\prime ~\approx~ \varepsilon\, H \,+ \ldots~,~~~
 \qquad H_d^\prime ~ \approx \varepsilon \,
i\sigma^2\,H^* \,+ \ldots~,
\eeq
where $\beta$ is a mixing angle and the ellipses represent terms at higher order in $\varepsilon$. In particular, we see that at leading order $H$ only has components in the original Higgs doublet. This means that the matter Yukawa couplings will have the same structure as in Split-SUSY at low energy. Furthermore, the presence of a light SM-like Higgs implies at first order in $\varepsilon$   
\begin{equation}
 B_\mu \simeq \sqrt{ (m_{H_u}^2 + \mu_0^2)(m_{H_d}^2 + \mu_0^2)} + \scr{O}(\varepsilon) \ .
\end{equation}

\subsection{Type II FSSM}

We present now a new model which realises the FSSM below $\MS$. The idea here is that only the higgsinos become fake in the low-energy theory. We shall refer to this as the type II FSSM (or FSSM-II for short). 

Since we do not have fake gauginos, the ultraviolet model building is much more conservative than the FSSM-I; in particular one does not have to appeal to Dirac gauginos. Instead, we just add two pairs of Higgs-like doublets, $\mathbf{H}_u^\prime, \mathbf{H}_d^\prime $ and $\mathbf{R}_u, \mathbf{R}_d$. Unification of the gauge couplings at one-loop above $\MS $ is recovered by adding two pairs of supermultiplets in the representations $(\mathbf{3}, \mathbf{1})_{1/3} \oplus  (\ov{\mathbf{3}}, \mathbf{1})_{-1/3}$. In total, we have therefore added two vector-like pairs of $\mathbf{5} + \ov{\mathbf{5}} $ of $SU(5)$. This should be reminiscent of gauge mediation scenarios, except that here the doublets mix with the Higgs fields.  

In order to create a split spectrum, we introduce an approximate R-symmetry with charges:
\begin{center}
\begin{tabular}{|c|c|}\hline
Superfields & R-charge \\\hline
$ \mathbf{H}_u,  \mathbf{H}_d$ & $0$ \\ \hline
$\mathbf{R}_u, \mathbf{R}_d$ & $2$ \\ \hline
$\mathbf{H_u^\prime},\mathbf{H_d^\prime}$ & $+1,- 1$  \\ \hline 
\end{tabular}\end{center}
Parametrising the breaking of this R-symmetry by a small parameter $\ehat$, the part of the superpotential containing the $\mu$ terms of the three Higgs-like multiplets is

\begin{eqnarray}
\label{superpotentialtype1}
W & \supset & \ehat^2 (\mu \mathbf{H}_u \, \mathbf{H}_d  + 
\mu_{H^\prime}  \mathbf{H}_u^\prime \, \mathbf{H}_d^\prime) \nn\\
& + & [\mu_{u} \mathbf{H}_u \,\mathbf{R}_d + \mu_{d} \mathbf{R}_u \, 
\mathbf{H}_d  ]   \nn\\
& + & \ehat   \mu_{fdr} \mathbf{R}_u \, \mathbf{H}_d^\prime + \ehat 
\mu_{df} \mathbf{H}^\prime_u \, \mathbf{H}_d + \ehat^3 \mu_{uf}  
\mathbf{H}_u \, \mathbf{H}^\prime_d . \nn
\end{eqnarray}
The R-charges have been chosen so that the mixing terms between $\mathbf{H}_{u,d}$ and $\mathbf{R}_{u,d}$ fields are unsuppressed. This allows the particles described mainly by $\mathbf{H}_{u,d}$ and $\mathbf{R}_{u,d}$ to have masses of order $\MS$, while $\mathbf{H}^\prime_{u,d}$ provide a pair of light F-higgsinos with a mass of ${\cal O}(\ehat^2 \MS)$. The Yukawa part of the superpotential is given by 
\begin{eqnarray}
W & \supset &
   [ Y_u \, \mathbf{U^c \, Q \cdot H_u}
- Y_d \, \mathbf{D^c \, Q \cdot H_d}
- Y_e \, \mathbf{E^c \, L \cdot H_d} ]\nn\\
  && + \ehat [
- Y_d \, \mathbf{D^c \, Q \cdot H^\prime_d}
- Y_e \, \mathbf{E^c \, L \cdot H^\prime_d} ] \nn
\end{eqnarray}
which allows a successful mass generation for the quarks and leptons, the SM-like Higgs obtained from fine-tuning at the electroweak scale must originate from the $\mathbf{H}_u$ and $\mathbf{H}_d$ multiplets. 

Imposing the R-symmetry on the soft terms leads to the suppression of the Majorana gauginos mass by $\ehat^2$ factors (this mechanism is similar to the usual Split SUSY one). In the term of the vector $ v_H\equiv (H_u, {H_d}^*,
H_u^\prime, H_d^{\prime\,*} , R_u , R_d^{*})$,  the Higgs mass matrix has the following hierarchy
\begin{align}
-\frac{1}{\MS^2}\mathcal{L}_{soft} ~~\supset ~~& v_H^\dagger \left(
\begin{array}{cccccc} \mathcal{O}(1) & \mathcal{O}(1) & 
\mathcal{O}(\ehat) &  \mathcal{O}(\ehat) & \mathcal{O}(\ehat^2) &
\mathcal{O}(\ehat^2)\\
  \mathcal{O}(1) & \mathcal{O}(1) &  \mathcal{O}(\ehat) &  
\mathcal{O}(\ehat) &  \mathcal{O}(\ehat^2) &
\mathcal{O}(\ehat^2)\\
\mathcal{O}(\ehat) & \mathcal{O}(\ehat) &  \mathcal{O}(1) & \mathcal{O}(1) &
   \mathcal{O}(\ehat )& \mathcal{O}(\ehat^3)\\
\mathcal{O}(\ehat) & \mathcal{O}(\ehat) &  \mathcal{O}(1) & \mathcal{O}(1) &
\mathcal{O}(\ehat) & \mathcal{O}(\ehat^3)\\
   \mathcal{O}(\ehat^2) & \mathcal{O}(\ehat^2) &
   \mathcal{O}(\ehat) & \mathcal{O}(\ehat) & \mathcal{O}(1) & 
\mathcal{O}(\ehat^4)  \\
   \mathcal{O}(\ehat^2) & \mathcal{O}(\ehat^2) &\mathcal{O}(\ehat^3) & 
\mathcal{O}(\ehat^3) & \mathcal{O}(\ehat^4)& \mathcal{O}(1)
\end{array} \right) v_H .
\end{align}

We can tune the SM-like Higgs from the scalar components of $\mathbf{H}_u$ and $\mathbf{H}_d$ to get 
\beq
H_u ~\approx~ \, \sin\beta\, H\, + \ldots~, 
~~~ \qquad H_d ~\approx~ \,
\cos\beta\,i\sigma^2\, H^* \,+ \dots~,
\eeq
and the other Higgs-like scalars only enters the linear combination with $\ehat$ suppression. The fine-tuning condition can therefore be applied on the $B_\mu$ term similarly, with the exception that the $\mu$ terms are not $\ehat$-suppressed compared to the soft masses, leading to
 \begin{equation}
 B_\mu \simeq \sqrt{ (m_{H_u}^2 + \mu_{u}^2)(m_{H_d}^2 + \mu_{d}^2)} + \scr{O}(\varepsilon) \ .
\end{equation}

The parameter $\ehat$ is here also fixed by the requirement that the gauginos obtain a mass at the TeV scale
\begin{align}
 \varepsilon = \scr{O}(\sqrt{\frac{ \text{TeV}}{\MS}}) .
\end{align}

\section{Unification and fine-tuning in Fake Split SUSY}
\label{SEC:RESULTS}

In \cite{FSSM1} the constraints on the FSSM from the bottom-up were mapped out under the most general assumptions of cosmology and UV completion. The remarkable result was that the scenario is consistent with \emph{any} supersymmetry-breaking scale. 
Here we would like to examine how robust this is once we take additional constraints into account: 
\begin{enumerate}
\item We shall assume that the universe has a standard cosmology, i.e. any hidden sector heavy particles decay well before dark matter freezes out -- since we are considering high SUSY scales this is typically the case. We then populate the dark matter abundance of the universe with the lightest neutral stable fermion in our model, or at least do not overpopulate (as in the case of underabundant dark matter the remainder could consist of axions or other hidden-sector particles). 
\item We shall consider the effect of the spectrum of the UV theory on the low energy result; in particular in \cite{FSSM1} $\tan \beta$ was taken as a free parameter but in general this is determined by the high-energy theory. 
\end{enumerate}

In \emph{(non-fake)} Split SUSY there is a known tension between the Higgs mass, unification and tuning $\tan \beta$ because the tuning requires
\begin{align}
\mathrm{det} \matw[m_{H_u}^2 + |\mu|^2, -B_\mu][-B_\mu, m_{H_d}^2 + |\mu|^2] \simeq 0  \quad \rightarrow\quad \tan \beta = \sqrt{\frac{m_{H_d}^2 + |\mu|^2}{m_{H_u}^2 + |\mu|^2}}.
\end{align}
Unification in Split SUSY requires $\mu$ to be $<10$ TeV at $1\sigma$ or $<100$ TeV at $2\sigma$ \cite{minisplit} and thus for much larger values of $\MS$ we would have $\tan \beta \approx \sqrt{\frac{m_{H_u}^2}{m_{H_d}^2}}$. For high values of $\MS$ to match the known value of the Higgs mass is is necessary to have a small $\tan \beta$; in  \cite{minisplit} it was found that the largest value of $\MS$ thus compatible with unification and the correct Higgs mass was $10^8$ GeV, and that required $\tan \beta =1$ -- if $\tan \beta = 2$ instead it becomes $10^6$ GeV  -- but a tuning of the Higgs soft masses to achieve such a value of $\tan \beta$ is not justifiable; just as in the MSSM the RGE running from any given mediation scale tends to drive $m_{H_u}^2 <0$ via
\begin{align}
16\pi^2 \frac{d}{d \log \mu} m_{H_u}^2 = 6 |y_t|^2 (m_{H_u^2} + m_Q^2 + m_U^2) + ...
\end{align} 
and this is exacerbated since the gaugino masses are much smaller than the scalar masses, so they cannot compensate. The conclusion is that without additional tuning $\tan \beta$ should be somewhat different from $1$, the SUSY scale should be low, and the amount of running from the scale at which the soft masses is generated cannot be too large (potentially problematic for gravity mediation). 

In fake Split SUSY, however, the situation is rather different although the details depend upon the high-energy theory: 

\begin{itemize}
\item In the FSSM-I, we have
\begin{align}
\mathrm{det} \matw[m_{H_u}^2 + |\mu_0|^2, -B_\mu][-B_\mu, m_{H_d}^2 + |\mu_0|^2] \simeq 0  \quad \rightarrow\quad \tan \beta = \sqrt{\frac{m_{H_d}^2 + |\mu_0|^2}{m_{H_u}^2 + |\mu_0|^2}}
\end{align}
as above but now unification only requires the \emph{fake-higgsino} mass parameter $\mu $ to be small which differs from $\mu_0$. This means that provided $\mu_0$ is sufficiently large it is not important whether $m_{H_u}^2$ becomes negative; we will always have a stable vacuum solution, and generically $\tan \beta \sim \mathcal{O}(1)$.

In addition, there is no R-symmetry protecting the masses and thus the RGEs take on the full dependence:
\begin{align}
16\pi^2 \frac{d}{d \log \mu} m_{H_u}^2 \simeq 6 |y_t|^2 (m_{H_u^2} + m_Q^2 + m_U^2 + A_t^2) - 6 g_2^2 M_2 - 2 g_Y^2 M_1 + 2 g_Y^2 \mathrm{tr} (Y m^2)
\end{align}
where the trilinear mass $A_t$ and gaugino masses $M_{1,2}$ are not suppressed. These can reduce the tendency for $m_{H_u}^2$ to become tachyonic.

\item In the FSSM-II, we have instead an R-symmetry which protects the trilinear scalar masses and gaugino masses, and neglecting terms of $\mathcal{O}(\varepsilon)$ we have 
\begin{align}
\mathrm{det} \matw[m_{H_u}^2 + |\mu_u|^2, -B_\mu][-B_\mu, m_{H_d}^2 + |\mu_d|^2] \simeq 0  \quad \rightarrow\quad \tan \beta = \sqrt{\frac{m_{H_d}^2 + |\mu_d|^2}{m_{H_u}^2 + |\mu_d|^2}}.
\end{align}
As in the FSSM-I, since $\mu_{u,d} \sim M_S$ there is no incompatibility with unification and obtaining $\tan \beta \sim \mathcal{O}(1)$.
\end{itemize}

Therefore there should be no impediment from taking the soft masses to be generated at the unification scale. We shall in the following consider the predictions from a scenario where this is the case: we shall take a common scalar mass $m_0$, common gaugino mass $M_{1/2}$ and (in the FSSM-I) a common trilinear mass $A_0$ at that scale and investigate the consequences for the Higgs mass and dark matter.

\subsection{Higgs mass and unification}
\label{sec:mH}

\begin{figure}[!htbp]
\begin{center}
\includegraphics[height=0.45\textheight]{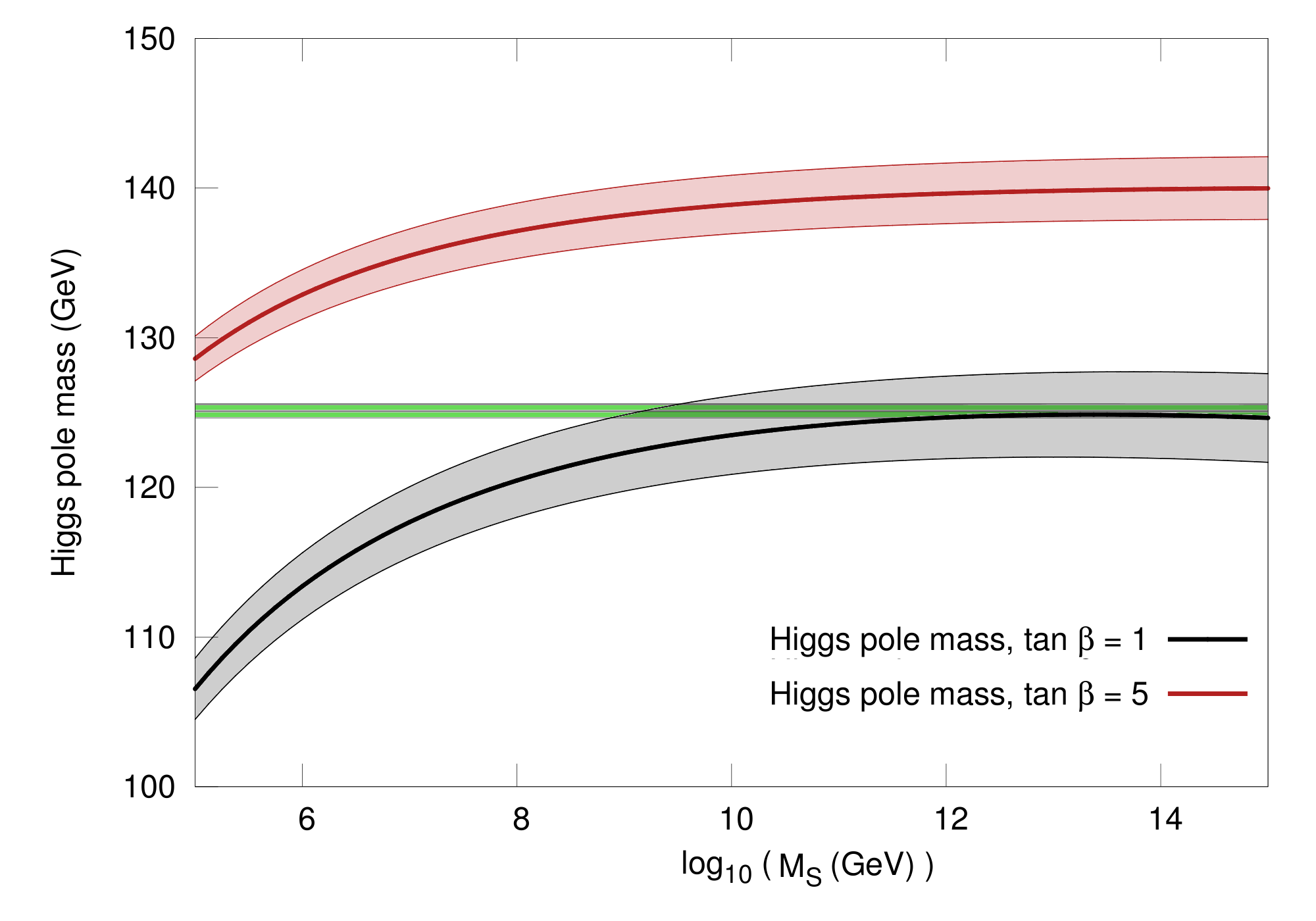}
\caption{  Higgs pole mass as a function of the SUSY scale, all parameters at the GUT scale have been set to be equal to the SUSY scale. The low energy spectrum is taken as $m_{\textrm{fg}} = 1$ TeV and $\mu_{f} = 1$ TeV. We consider a Non Universal Higgs Mass (NUHM) scenario in FSSM-II so that we fix directly $\tan \beta$ at $\MS$ to $1$ for the lower curve and $5.$ for the upper one. The shaded region gives the variation from a $2 \sigma$ variation in the top pole mass. The green band corresponds to the measured Higgs mass.}
\label{fig:mh_msusy}
\end{center}
\end{figure}

We have implemented the calculation of the spectrum of the FSSM at low energies based on high-energy boundary conditions in a code as described in Appendix~\ref{APP:implementation}. Here we wish to revisit the prediction of the Higgs mass from \cite{FSSM1} once we impose unified boundary conditions in the UV. 
The Higgs mass as a function of  $\MS$ is shown in Figure~\ref{fig:mh_msusy} (where all heavy mass parameters have been taken to be equal to the SUSY scale). 
The slightly higher Higgs mass than \cite{FSSM1} arises because the running from the GUT scale produces somewhat heavier gluinos; Figure 4 of~\cite{FSSM1} describes this effect. In the plot, it is useful to note that the curves exhibit plateaux so that by choosing the right value of $\tan \beta$ between $1$ and $5$ we can reproduce the desired Higgs mass for any SUSY scale up to the GUT scale.

\begin{figure}[!htbp]
\begin{center}

\includegraphics[height=0.38\textheight]{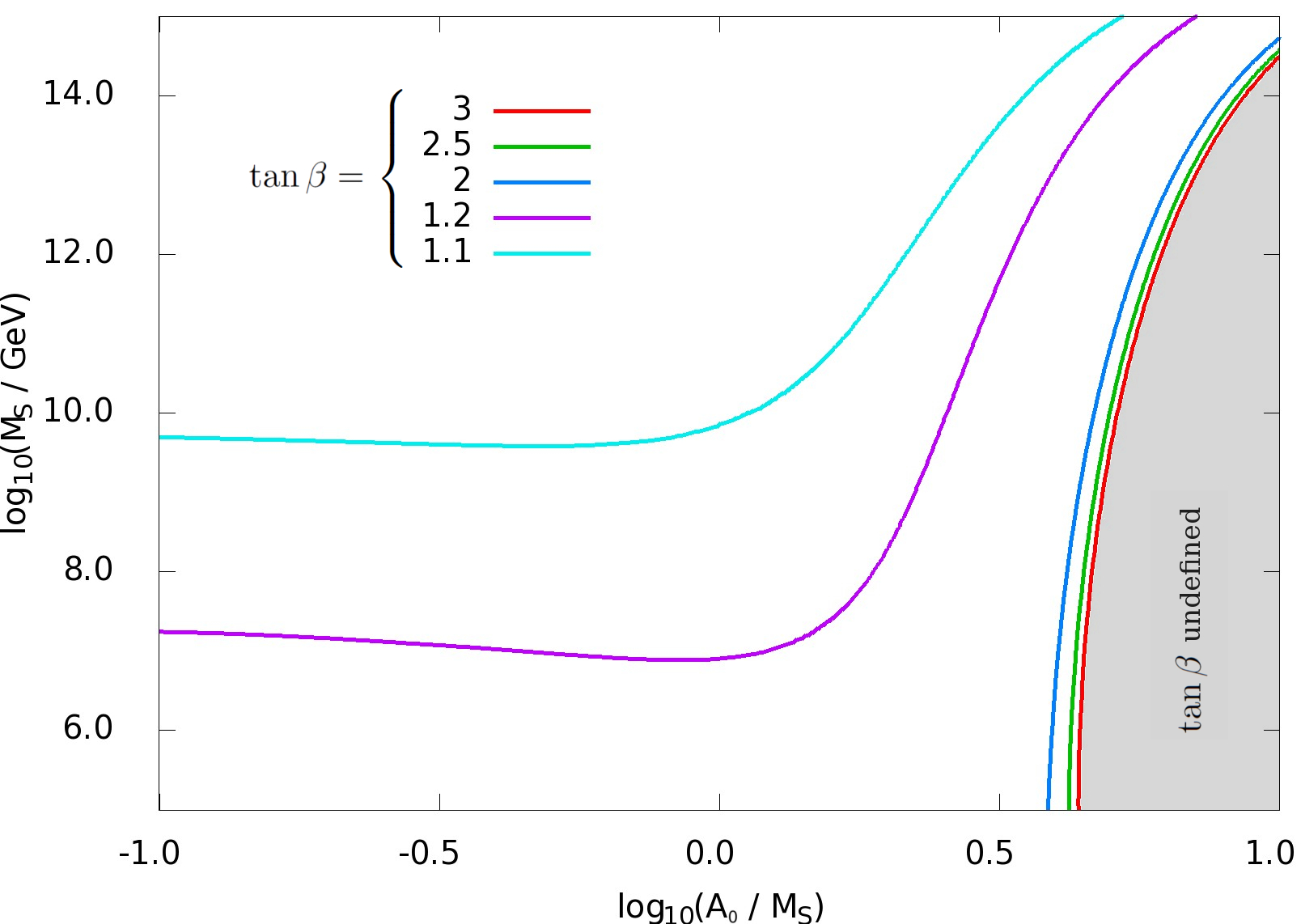}
\caption{  Contours of the value of $\tan \beta  = \sqrt{\frac{m_{H_d}^2 + \mu_0^2}{m_{H_u}^2 + \mu_0^2}} $ found to match the observed Higgs mass in the FSSM-I varying the scalar unification mass $m_0$ and trilinear mass $A_0$. }
\label{fig:tanb_at_m0}
\end{center}
\end{figure}

\begin{figure}[!htbp]
\begin{center}
\includegraphics[height=0.38\textheight]{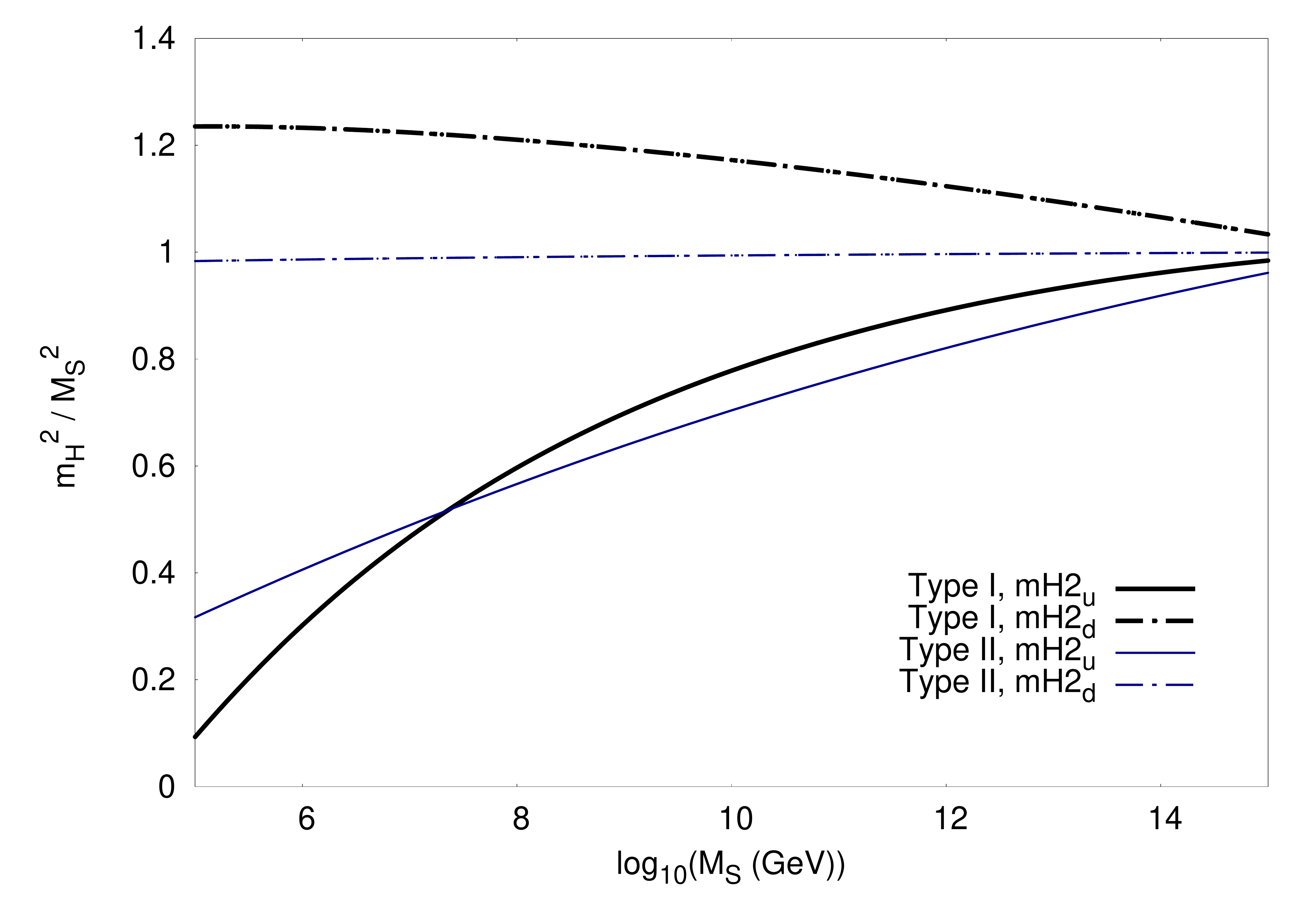}
\caption{ Running masses $m^2_{H_u}/M_S^2$ and $m^2_{H_d}/M_S^2$ in the FSSM-I (bold lines) and the FSSM-II (normal line) at the SUSY scale as a function of the $M_S$. All UV parameters are set to be equal to the SUSY scale. }
\label{fig:mh2_ms7}
\end{center}
\end{figure}

If we suppose unification of the Higgs masses at the GUT scale (so that $m^2_{H_u} \simeq m^2_{H_d}$ and $ \mu_{u} \simeq \mu_{d} $), then $\tan \beta $,  all parameters in (\ref{tanbeta}) are of the same order, and we predict that generically the value of $\tan \beta $ is close to $1$. This can be seen in Figure~\ref{fig:tanb_at_m0} where we have plotted  $\tan \beta $ in the FSSM-I as a function of unified SUSY-breaking scalar mass $m_0$ and the A-term at the GUT scale $A_0$. We see that in most of the parameter space  $\tan \beta $ is  between $1$ and $1.4$. The increase in the right part of the plot show that for a larger value of $A_0$, $m_{H_u}^2 + \mu_0^2$ can run close to zero. In principle, by varying $m_0$ and $A_0$ in the FSSM-I we can find values of  $\tan \beta  > 2$,  potentially allowing values of the SUSY scale lower than $10^9$ GeV without requiring a breaking of the universality of the soft masses at the GUT scale. Figure~\ref{fig:mh2_ms7} illustrates the effect of the running of the Higgs soft masses: in both the FSSM-I and the FSSM-II the renormalisation group evolution does not greatly separate these masses leading to a $\tan \beta \simeq \scr{O}(1) $. Note that the longer the running above $\MS$, the higher the predicted $\tan \beta$, which in turn raises the Higgs mass at tree level. Hence for small values of $\MS$ it is natural to have larger values of $\tan \beta$, and for larger $\MS$ we expect $\tan \beta \sim 1$, both compatible in this model with the observed Higgs mass.

\begin{figure}[!htbp]
\begin{center}
\includegraphics[height=0.38\textheight]{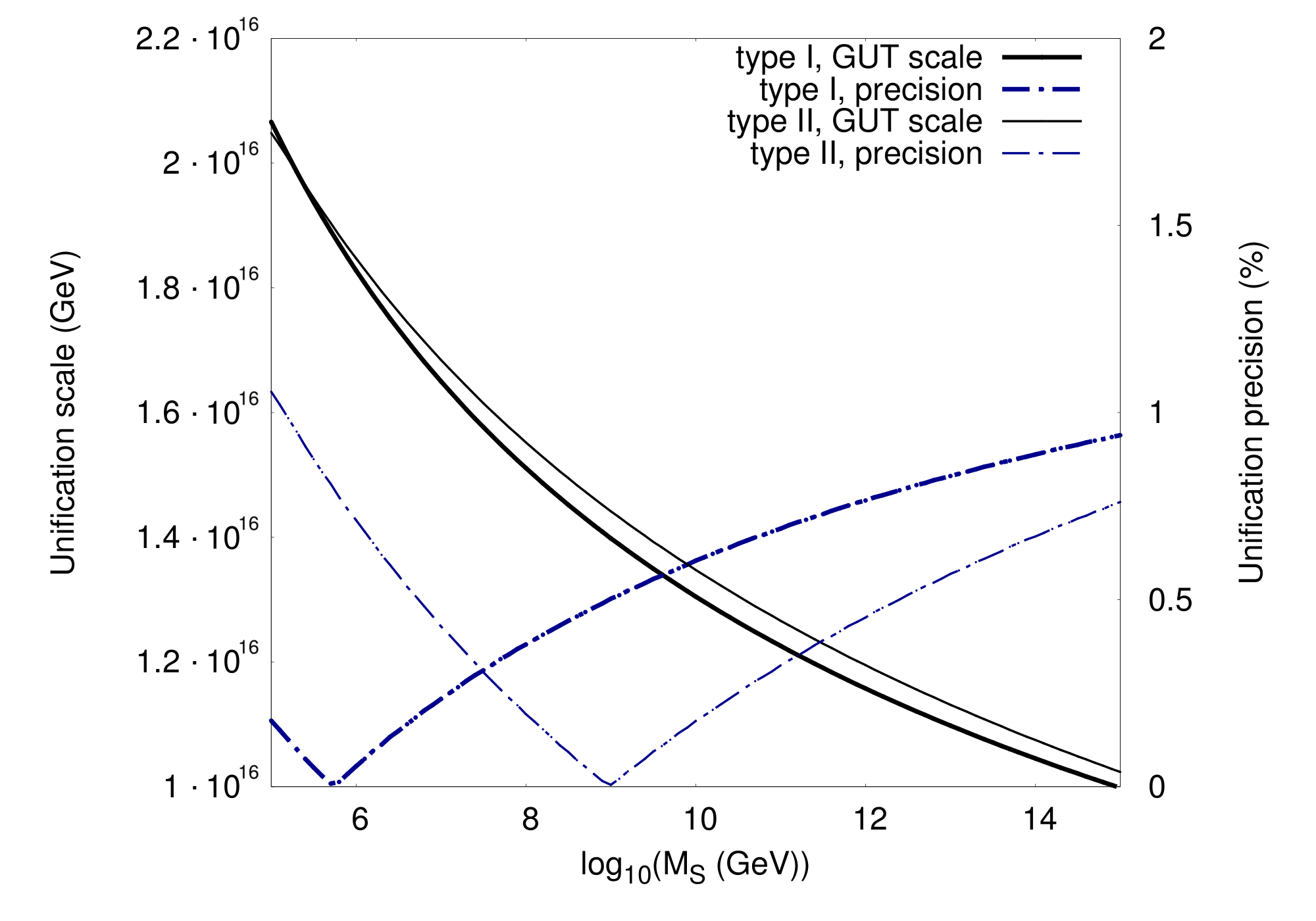}
\caption{ Evolution of the unification scale as well as the precision of the unification ($|g_1- g_3|/g_3$ in percent at the point where $g_1$ and $g_2$ unify) as a function of the SUSY scale $M_S$. All UV parameters are set to be equal to the SUSY scale. }
\label{fig:unif_msusy}
\end{center}
\end{figure}

As we discussed above, unification in both models is ensured at one-loop. At two-loops it is also well preserved, as can be seen from Figure~\ref{fig:unif_msusy} where we have plotted the unification scale as a function of the SUSY scale $\MS$, along with $|g_1- g_3|/g_3$ at the unification scale of $g_1$ and $g_2$. A percent level unification can be obtained for all $\MS$ for FSSM-I and above $10^7$ GeV for FSSM-II. 
The unification scale itself remains of the order of $ 10^{16}$ GeV. 

It should be added that as was noted in~\cite{FSSM1} that for certain regions of the parameter space the Higgs quartic coupling can become (slightly) negative during its running between the electroweak and the SUSY scale. This feature appears however for SUSY scales above $10^{13}$ GeV. 

\subsection{Dark matter and cosmology}

In this subsection, we investigate the consequences of assuming a thermal history of the Universe: avoiding an overly long-lived gluino destroying BBN (or even surviving to be ruled out today); requiring the neutralino LSP to provide a dark matter candidate with the correct relic density (or at least not an overdensity);  and escaping  current direct detection limits. 

At low energies, the non-SM fields in the FSSM are organised into a set of neutral fermions -- neutralinos --  and charged ones -- charginos. In the basis
$(\tilde{B}^\prime,\tilde{W}^{\prime\,0}, \tilde{H'_d}^0,
\tilde{H'_u}^0)$ the neutralino mass matrix is
\begin{equation}
\mathcal{M}_{\chi^0} = \left(\begin{array}{c c c c}
 \mtBp   & 0       &  \displaystyle -  \frac{\tilde{g}_{1d} v}{\sqrt{2}} & \displaystyle  \frac{\tilde{g}_{1u} v}{\sqrt{2}} 
\\
0     & \mtWp  &  \displaystyle   \frac{\tilde{g}_{2d} v}{\sqrt{2}} & \displaystyle -  \frac{\tilde{g}_{2u} v}{\sqrt{2}} \\
\displaystyle-  \frac{\tilde{g}_{1d} v}{\sqrt{2}} & \displaystyle \frac{\tilde{g}_{2d} v}{\sqrt{2}}  & 0  & -\mu \\
 \displaystyle \frac{\tilde{g}_{1u} v}{\sqrt{2}} & \displaystyle -  \frac{\tilde{g}_{2u} v}{\sqrt{2}}  & -\mu  & 0 \\
\end{array}\right)~.
\label{diracgauginos_NeutralinoMassarray2}
\end{equation}

We can express the chargino mass matrix involving the F-higgsinos $\tilde{H'}^+$,
$\tilde{H'}^-$ and the charged (F-)gauginos $\tilde{W^\prime}^\pm$ in
the form
\begin{equation}
- (v^-)^T\mathcal{M}_{\chi^\pm}  v^+ ~+~ {\rm h.c.}~,
\label{diracgauginos_CharginoMassLagrangian}
\end{equation}
where we have adopted the basis $v^+ =
(\tilde{W^\prime}^+,\tilde{H'}^+_u)$, $v^- =
(\tilde{W^\prime}^-,\tilde{H'}^-_d)$. This reads
\begin{equation}
\mathcal{M}_{\chi^\pm}~ =~
\left(\begin{array}{c c}
\mtWp  &  \ \tilde{g}_{2u} v \\
  \tilde{g}_{2u} v  & \mu \\
\end{array}\right) .
\label{diracgauginos_CharginoMassarray2}
\end{equation}
Here the crucial difference to Split SUSY is the the suppression of  the F-higgsino Yukawa couplings $\tilde{g}_{iu,d} $ (by $\varepsilon$ for the FSSM-II and $\varepsilon^2$ for the FSSM-I), which results in rather different dark matter phenomenology. We will consider the standard three possible scenarios for a viable Dark Matter candidates:
\begin{itemize}
 \item Scenario $\HDM$: F-higgsino LSP.
 \item Scenario $\WDM$: (F-)Wino LSP.
 \item Scenario $\BHDM$: a mixed F-Bino/F-higgsino LSP, with a small splitting.
\end{itemize}

Notice that a priori, one can also have a mixed Bino/Wino dark matter which gives the correct relic density. But since we expect generically that the gaugino mass hierarchy is fixed by the chosen mechanism of supersymmetry breaking, one does not have the freedom to tune the (F-)Bino / (F-)Wino mass ratio as can the (F-)Bino and the F-higgsinos masses in the scenario $\BHDM$. We shall not discuss here such a scenario.

In the setup of  $\WDM$, since the RG running would naturally induced a Bino LSP, one has to consider non-universal gaugino masses (NUGM) at the GUT scale.   For practical purposes, we will consider unification at $M_{GUT}$ between the Wino and gluino masses but suppose that the SUSY breaking mechanism induces a larger Bino mass. The latter becomes an extra parameter which has no impact on the Higgs mass and on the Dark Matter constraint, as long as it is heavy enough not to be the LSP. In the following, when dealing with scenario $\WDM$, we take $M_1=10$ TeV at the GUT scale, which translates into a Bino of roughly $5$ TeV at the electroweak scale. 

Finally the scenario $\BHDM$ relies on co-annihilation between the higgsinos and Binos to avoid overproduction of  the latter. This implies that the Bino mass must be chosen precisely to reproduce the correct relic density. Evaluating fine-tuning from the simplest definition:
\begin{equation}
\Delta =  \frac{\partial \Omega h^2}{\partial  m_{\tilde{B}}} \frac{m_{\tilde{B}}}{\Omega h^2} \ ,
\end{equation}
we found $\Delta \sim 20-40$ in the scenario $\BHDM$ (depending on $\MS$ and on wether or not one consider FSSM-I or FSSM-II), while we have $\Delta \sim 1$ in the scenarios $\HDM$ and $\Delta \sim 2$ in $\WDM$, indicating the this scenarios is ten times more fine-tuned than the two others. It however offers  other virtues, such as avoiding the constraints from direct detection which apply for $\HDM$. 

In order to compute the relic density, we have used routines from the code\\
\micrOMEGAs~\cite{belanger_micromegas_3:_2014}. This  is supplemented by the constraints from the gluino life-time and   from direct detection experiments which become relevant when our candidate is an almost Dirac fermion as it can happen with  F-higgsino Dark Matter.

\subsubsection{Relic density}
\label{sec:RelicDensity}

The LSP abundances are governed  mainly by gauge interactions that are the same for true and fake gauginos/higgsinos. The  suppressed Yukawa couplings are expected to play a minor role. In that case, one can use the standard expressions \cite{arkani-hamed_well-tempered_2006} to obtain a rough estimate
\begin{align}
 \Omega_{\tilde{W}} h^2 = 0.13 \left( \frac{M_2}{2.5 \text{ TeV}} \right)^2 \ , 
\end{align}
for Wino-like DM and 
\begin{align}
 \Omega_{\tilde{H}} h^2 = 0.10 \left( \frac{\mu}{1 \text{ TeV}}\right)^2 \ ,
\end{align}
for higgsino-like dark matter.

We have used the public code \micrOMEGAs~\cite{belanger_micromegas_3:_2014} to compute the relic density in the three scenarios described above. We used \SARAH~\cite{staub_superpotential_2010} to generate the CalcHep file which was taken as an input by \micrOMEGAs. We take for the relic density the Planck 2015 value~\cite{ade_planck_2015} $\Omega h^2 = 0.1188 \pm 0.0010$; clearly the \emph{theoretical} uncertainty stemming from higher-order corrections is many times larger than this -- the contours could potentially move by potentially as much as $50\%$. However, we do not show this uncertainty in the plots because it is difficult to estimate, and because the important point is the relationship between the parameters. The reader should just be wary of taking our numbers as absolute.

In scenarios $\HDM$ and $\WDM$, our results are fully consistent with the previous approximate formulas. In order to recover the correct relic density at $3 \sigma$, we need to have an  F-higgsino pole mass between $1110$ GeV and $1140$ GeV or a (F-)Wino pole mass between $2390$ GeV and $2450$ GeV.

\begin{figure}[!htbp]
\begin{center}
\includegraphics[height=0.50\textheight]{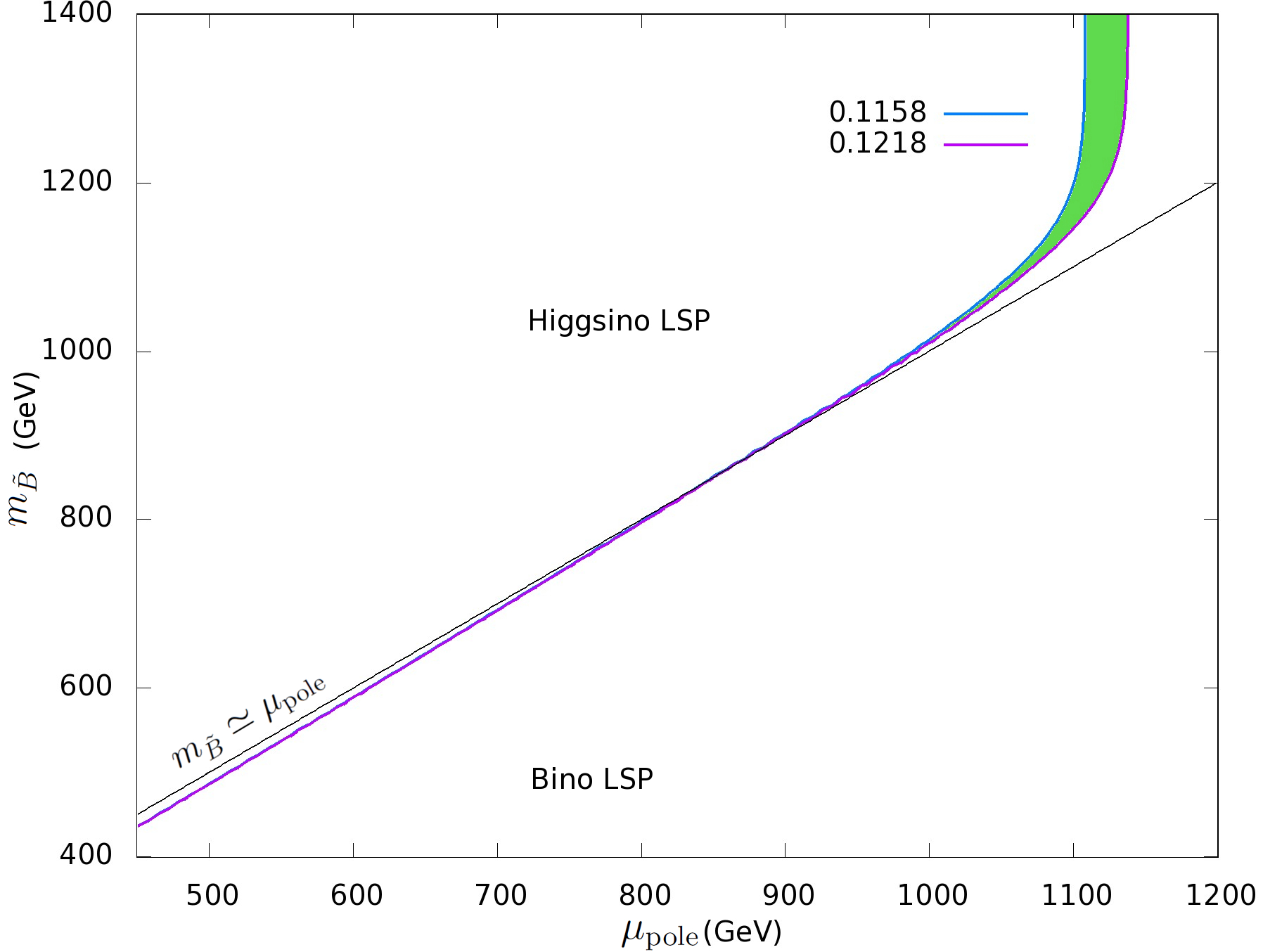}
\caption{ Visualisation of the constraints coming from gluino life-time, from the requirement of a $125$ GeV Higgs pole mass, and from obtaining the correct relic density. We furthermore represent the separation (Black diagonal line) between a Bino LSP and a Higgsino LSP We use a $\mu_\text{pole}$ - $m_{\tilde{B}}$ plane, where $m_{\tilde{B}}$ is the Bino pole mass and $\mu_\text{pole}$ is the Fake Higgsinos pole mass. The SUSY scale $M_S$ has been chosen at $10^{10}$ GeV. Calculations has been done in the FSSM-II. }
\label{fig:const_mixed}
\end{center}
\end{figure}

In general in the FSSM the mixing between the Bino and the higgsino will be very small; the mixing is controlled by $\frac{\tilde{g}_{1u,d} v}{|\mu| - |M_1|}$. For example, if we take $\MS = 10^{9}$ GeV then $\varepsilon \sim 10^{-3}$ so for $(|\mu| - m_{\tilde{B}}) \sim v$ we have mixing in the FSSM-II of $\sim 10^{-3}$ and in the FSSM-I of $\sim 10^{-6}$. Since the Bino cannot annihilate except through mixing, in the $\BHDM$ scenario we therefore require coannihilation to obtain the correct relic density. However, differently to other SUSY scenarios, when we have coannihilation so that $|\mu_{\mathrm{pole}} - m_{\tilde{B}}| \lesssim T_{f}, $ the temperature at freezeout, the mixing is in general still very small: since as usual $T_f \sim m/20 \sim \mathcal{O}(10)$ GeV for $m$ the LSP mass, the enhancement of the mixing is only $\mathcal{O}(10)$ -- which for small values of $\varepsilon$ still leads to negligible mixing of the Bino/F-higgsino. Only when $\MS$ is rather low and in the FSSM-II, or in the case of very small mass differences, smaller than  that required to allow coannihilation, will we find appreciable mixing. 

To be more explicit, consider that pure higgsinos have an annihilation cross-section given by
\begin{align}
\bra \sigma_{\tilde{H}\tilde{H}} v\ket \simeq & \frac{g^4}{512 \pi \mu^2} (21 + 3 \tan^2 \theta_W + 11 \tan^4 \theta_W)
\end{align}
and their interactions freeze out at the typical temperature of $T_f\sim \mu/x_f$ where $x_f \equiv m/T_f \simeq 25$. So if the Bino has a similar mass but weakly mixes, let us approximate the 
ratio $\Gamma/H \equiv n \bra \sigma v \ket/H $ for processes involving it near the freezeout temperature and put $m \sim M_W$:
\begin{align}
\frac{\Gamma (\tilde{B} + \tilde{H} \rightarrow\mathrm{SM\ fermions})}{H} &\sim \frac{\tilde{g}_{iu,d}^2}{M_W^2} \frac{(mT)^{3/2} e^{-m/T} }{1.66\sqrt{g_*} T^2/M_P} \sim 10^{4} \times \tilde{g}_{iu,d}^2  \nn\\
\frac{\Gamma (\tilde{B} + \mathrm{SM} \rightarrow \tilde{H} + \mathrm{SM})}{H} &\sim \frac{\tilde{g}_{iu,d}^2}{M_W^2} \frac{T^3 }{T^2/M_P} \sim 10^{16}\times \tilde{g}_{iu,d}^2
\end{align}
so the first process is always frozen out well before the higgsino interactions, but the second will remain important for $\MS \lesssim 10^{11}$ GeV in the FSSM-I and for \emph{any} value of $\MS$ up to the Planck scale in the FSSM-II. This means that the Bino remains thermalised even if its annihilations are ineffective. We can therefore calculate the relic density rather straightforwardly following \cite{Griest:1990kh}: defining $\Delta_i \equiv \frac{m_i - m}{m} $ and
\begin{align}
r_i \equiv n_i^{eq}/n^{eq} =& \frac{g_i (1+ \Delta_i)^{3/2} \exp (- x \Delta_i)}{\sum_{i=1}^N g_i (1+ \Delta_i)^{3/2} \exp (- x \Delta_i)}  \nn\\
\bra \sigma_{eff} v \ket \equiv& \sum_{i,j}^N r_i r_j \bra\sigma_{ij} v\ket
\end{align}
 we have
 \begin{align}
 \Omega h^2 \simeq& \frac{8.7 \times 10^{-11} \mathrm{GeV}^{-2}}{\sqrt{g_*} \int_{x_f}^\infty dx x^{-2} \bra \sigma_{eff} v \ket}.
 \end{align}
The integral over temperatures \emph{after} the freezeout (in the denominator) can be important as there can be a significant reduction of the dark matter density. 

Let us define $\Omega_c h^2 (= 0.1188)$ as the observed dark matter density fraction, and $\mu_c$ the value of $\mu$ that matches this for a pure higgsino. Then for our case we can approximate 
\begin{align}
\bra \sigma_{eff} v \ket = r_{\tilde{H}}^2 \bra \sigma_{\tilde{H}\tilde{H}} v\ket \simeq  r_{\tilde{H}}^2 \times 8.7 \times10^{-11} x_f/\sqrt{g_*} \times \left( \frac{\mu_c}{\mu} \right)^2 \times \frac{1}{\Omega_c h^2}
\end{align}
so that
\begin{align}
 \frac{\Omega h^2}{\Omega_c h^2}\simeq& \left(\frac{\mu}{\mu_c}\right)^2 \frac{1}{x_f \int_{x_f}^\infty dx \ r_{\tilde{H}}^2/x^2}.
 \label{eq:OmegaApprox}\end{align}
Therefore if we plot the contour matching the relic density in the Bino-higgsino mass plane, as we have done in Figure~\ref{fig:const_mixed}, we are plotting the contour of the right hand side of the above equal to one. We find in the FSSM, since we shall typically require $\Delta_i \ll 1$, that we can well approximate 
\begin{align}
r_{\tilde{H}} \simeq \bigg(1 +\frac{1}{4} \exp\big[-x \left( \frac{m_{\tilde{B}} - m_{\tilde{H}}}{m_{\tilde{H}}}\right)\big]\bigg)^{-1} .
\end{align}
The immediate observation is that when $m_{\tilde{B}}= m_{\tilde{H}}$ we have $r_{\tilde{H}} = 4/5$ and so we require $\mu = \frac{4}{5}\mu_c$; on figure \ref{fig:const_mixed} we see that the curves cross at $900$ GeV which is exactly four fifths of $1125$ GeV, the critical value for a pure higgsino. This crossing point can be of importance, since F-higgsino dark matter is a perfect example of inelastic dark matter and therefore direct detection experiments can be sensitive to it. Numerically evaluating equation (\ref{eq:OmegaApprox}) then gives a curve in excellent agreement with the results of \micrOMEGAs. For a Bino LSP we find a linear approximation to fit rather well in the range of values considered $m_{\tilde{B}} \simeq \mu_{\rm pole} - (4 \mu_c/5 - \mu_{\rm pole})/x_f$, i.e. the mass difference required is of order $T_f$.

\subsubsection{Direct detection and inelastic scattering }

We have computed the conventional direct detection constraints for our model and found that, when the dark matter can be treated as a Majorana particle, due to the highly suppressed Higgs/(F-)gaugino/F-higgsino interactions, they barely restrict the parameter space. However, since those same interactions determine the splitting between the F-higgsino mass eigenstates, when it is small enough the fake higgsinos can be treated as a Dirac fermion. In that case one can have vector-vector couplings with nucleons through the exchange of a $Z$ boson, leading to inelastic scattering. The spin-independent cross-section implied by this process is so large that direct detection experiments have already ruled these out by many orders of magnitude. This effect has been studied in~\cite{nagata_higgsino_2015} where they find that the XENON100~\cite{aprile_implications_2011} and LUX~\cite{akerib_first_2014} experiments constrained the splitting to be larger than $210$ keV for a $1$ TeV higgsino LSP. We will consider below a conservative bound of $300$ keV for the splitting.

Given the mass matrices for neutralino~(\ref{diracgauginos_NeutralinoMassarray2}) the splitting $\delta$ between the two higgsinos can be estimated as :
\begin{align}
 \delta \simeq -\frac{v^2}{4} \left[ \frac{(\gtdp + \gtup)^2}{M_1-\mu}+\frac{(\gtd + \gtu)^2}{M_2-\mu}+\frac{(\gtdp - \gtup)^2}{M_1+\mu}+\frac{(\gtd - \gtu)^2}{M_2+\mu} \right] \ .
\end{align}
This analytic formula agrees with the numerical mass difference between the two higgsinos pole masses at a few percent level accuracy when estimated using $\msbar$ running parameter at the electroweak scale. This gives
\begin{align}
\label{eq:approxsplit}
 \delta & \simeq  \begin{cases}
200 \ \text{keV} \cdot \scr{O}(1) \cdot \displaystyle \left( \frac{  400 \text{ TeV} }{ \MS}\right)^2   \left( \frac{  m_{fg} }{ 4 \text{ TeV}}\right) \quad \textrm{ for the FSSM-I}  \\
200 \ \text{keV} \cdot \scr{O}(1) \cdot \displaystyle \left( \frac{  10^7 \text{ GeV} }{ \MS}\right) \left(\frac{\mu }{ 1 \text{ TeV}}\right) \left( \frac{  4 \text{ TeV} }{ m_{fg}}\right) \quad \textrm{ for the FSSM-II}  \ ,
\end{cases} 
\end{align}

where $m_{fg}$ gives the typical scale of the F-gaugino masses. The extra $\scr{O}(1)$ terms come from the uncertainty on the precise suppression of the $\gtu, \gtd, \gtup$ and $ \gtdp$ couplings. We see that for F-gauginos of several TeV and for a $\mu$ term around $1$ TeV (as required from relic density constraints), the SUSY scale $\MS$ is bounded below roughly $5 \cdot 10^8$ GeV for the FSSM-II and $5 \cdot 10^6$ GeV for the FSSM-I if the $\scr{O}(1)$ is taken to be $10$. The constraints are far more stringent than in Split SUSY because of the extra-suppression in $\varepsilon^2$ for the FSSM-I and in  $\varepsilon$ for the FSSM-II.

\subsubsection{The (F-)gluino lifetime}

\begin{figure}[!htbp]
\begin{center}
\includegraphics[height=0.50\textheight]{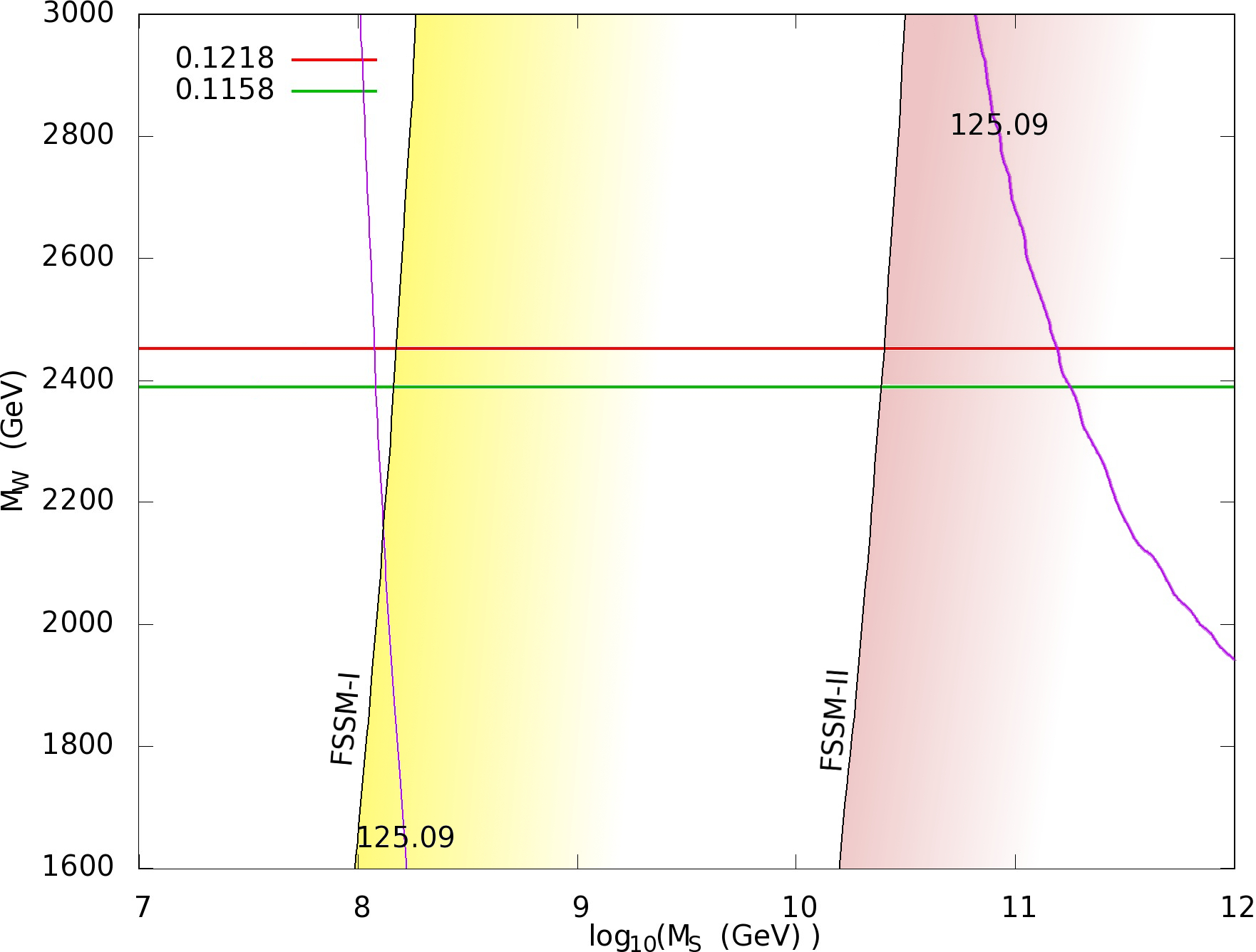}
\caption{ Visualisation of the constraints coming from gluino lifetime, from the requirement of a $125$ GeV Higgs pole mass, and from obtaining the correct relic density in scneario $\WDM$. We use a $M_S$ - $M_{\textrm{W}}$ plane, where $M_W$ is the Wino pole mass. The yellow color gradient indicate the area excluded with gluino life-time bigger than $100$ s in FSSM-I. The red color gradient is the area for the FSSM-II. The bold purple line gives $125$-GeV Higgs for $M_t = 173.34$, the slimmer one is the $125$-Gev Higgs for a $2 \sigma$ variation in $M_t$ }
\label{fig:const_nugm}
\end{center}
\end{figure}

In the FSSM-I, fake gluinos are even more long-lived than gluinos in usual Split Supersymmetry~(\cite{gambino_gluino_2005}, \cite{dudas_flavour_2013}). Indeed, the decay of F-gluinos to the lightest F-neutralino must proceed via mixing with the usual gluinos in order to have couplings to sfermions. And since the mixing is suppressed by factors of $\epsilon$, the overall F-gluino lifetime in the FSSM-I is therefore enhanced by a factor of $\epsilon^{-4} \simeq \frac{M_S^2}{m^2_{fg}}$.
\begin{align}
 \tau_{\tilde g'}  ~\sim~&
  4 \ \text{sec}\times\left(\frac{M_S}{10^7\text{GeV}}\right)^6
\times\left(\frac{1     \ \text{TeV}}{m_{fg}}\right)^7~.
\end{align} 
Since the gauginos are not fake in the FSSM-II, this enhancement does not occur and one is left instead  with the Split SUSY gluino life-time
\begin{align}
 \tau_{\tilde g'}  ~\sim~&
  4 \ \text{sec}\times\left(\frac{M_S}{10^9\text{GeV}}\right)^4
\times\left(\frac{1     \ \text{TeV}}{m_{fg}}\right)^5~.
\end{align} 
Constraints from Big Bang Nucleosynthesis (BBN) limit this lifetime to be below $100$s if one relies on a standard cosmology~\cite{arvanitaki_limits_2005}. A much longer lifetime gluino is constrained from the CMB spectrum, the gamma-ray background or even heavy-isotope searches when the gluino is stable at the scale of the age of the universe. As discussed in~\cite{FSSM1} they give very strong bounds on $\MS$ for a standard thermal history of the universe.

Overall, the effect of the previous formulas with our values for the pole masses can be visualised in Figure~\ref{fig:const_nugm} where we chose a Wino dark matter. We see that since the Wino pole masses must be quite heavy in order to get the correct relic density, the gluino pole mass ends up in the several TeV regime, reducing slightly the gluino lifetime. In $\WDM$ scenarios, the (F-)gluino lifetime gives an upper bound on the possible $\MS$ of $10^8 \text{ GeV}$ for the FSSM-I and of $10^{10} \text{ GeV}$ for the FSSM-II. One should not forget that the (F-)gluino pole mass is here obtained by supposing unification of the (F-)Wino and (F-)gluino masses at the GUT scale. These bounds should therefore be modified according to the previous formulas if one considers a particular SUSY breaking setup with a given ratio between (F-)gaugino masses.

\subsubsection{Summary of the cosmological constraints}

The direct detection for inelastic Dark Matter, the relic density, and the constraint on gluino life-time,  have set bounds on four parameters of our model: the F-higgsino pole masses $\mu_\text{pole}$, the (F-)Bino pole mass $m_{\tilde{B}}$, the (F-)Wino pole masses $m_{\tilde{W}}$ and the SUSY scale $\MS$. 

Even though some constraints depend non-trivially of several of these parameters, one can deduce from the previous analysis rough windows for each parameter in three Dark Matter scenarios we have studied. These windows are summarised in Table~\ref{table:constraints}.

\begin{table}[h]
\centering
\begin{footnotesize}
\begin{tabular}{|l||p{3.5cm}|l|l|}
  \hline
 \textbf{DM type} & \textbf{Inelastic scattering} & \textbf{Relic density} & \textbf{Gluino lifetime} \\[0.5em]  \hline  
\rule{0pt}{3ex}  \textbf{$\mathbf{\WDM}$}    & None &  $m_{\tilde{W}} \subset [2390,2450]$ GeV  & \multirow{3}{3.7cm}{For multi-TeV  gluinos  \\ $ \begin{cases}
          M_S \lesssim 5 \cdot 10^{8} \text{GeV} \\  \text{(for FSSM-I)} \\ \\
	  M_S \lesssim  2 \cdot 10^{10} \text{GeV} \\  \text{(for FSSM-II)}
         \end{cases} $ } \\[0.8em] \cline{1-3}
\rule{0pt}{3ex}   \textbf{$\mathbf{\BHDM}$} & $\mu_\text{pole}  \lesssim 900$ GeV& $m_{\tilde{B}} \simeq \mu_\text{pole}  - (900 - \mu_{\rm pole})/x_f$&  \\[0.3em] \cline{1-3} 
\rule{0pt}{3ex}  \textbf{$\mathbf{\HDM}$} &  $ \begin{cases}
          M_S \lesssim 5 \cdot 10^{6} \text{GeV} \\  \text{(for FSSM-I)} \\ \\
	  M_S \lesssim  10^{8}  \text{GeV}  \\  \text{(for FSSM-II)}
         \end{cases} $
 & $\mu_\text{pole} \subset [1110,1140] $ GeV   &     \\[1em]
 \hline
\end{tabular}
\end{footnotesize}
\caption{ \footnotesize Approximate constraints on the SUSY scale and on pole masses for the Dark matter candidates. We impose a splitting between fake Higgsinos bigger than $300$ keV to avoid direct detection through inelastic scattering, we require a gluino life-time smaller than $100$ s to avoid hampering BBN and finally constrain the relic density (calculated at tree-level in \micrOMEGAs) to be $ \Omega h^2 \subset [0.1158 ,0.1218 ]$. When considering constraints on $M_S$, gaugino masses were taken in the multi-TeV range. }
\label{table:constraints}
\end{table}

If we take $\tan\beta = 1$,  the Higgs mass gives a lower bound on the SUSY scale $M_S \gtrsim 5 \cdot 10^8 $ GeV, which in the FSSM-I is in tension with the gluino lifetime.  We  see from Table~\ref{table:constraints} that the $\HDM$ scenario is also almost ruled out by direct detection constraints depending on the precise suppression of $\gtu, \gtd, \gtup$ and $ \gtdp$, so we should predict that for such a value of $\tan \beta $ we should have a mixed Bino-higgsino dark matter candidate if the gaugino masses unify and in the FSSM-II only. 

The constraints from dark matter may present an upper bound on $\MS$ if we are unwilling to accept a coincidence of a few GeV between $\mu$ and the Bino mass, since in the $\HDM$ case even an underabundance of dark matter would be ruled out if the mass splitting is too small. However, if we would like to reach the bound on $\MS$ from the gluino lifetime without changing the cosmology of the universe, there are two possibilities:
\begin{itemize}
\item Introduce some R-parity violation so that our LSP decays. Then the dark matter should consist of axions. 
\item In the FSSM-II, we could consider a gravitino LSP. As discussed in \cite{FSSM1,Goodsell:2014dia}, for the FSSM-I this does not help. However, in the FSSM-II the gaugino decays to the gravitino are unsuppressed, so if we have a Bino LSP 
\begin{align}
\tau_{\tilde{B}} =& \frac{16\pi F^2}{m_{\tilde{B}}^5} \simeq 1\ \mathrm{s} \times \left( \frac{\MS}{10^{9}\ \mathrm{GeV}} \right)^4 \times \left( \frac{500\ \mathrm{GeV}}{m_{\tilde{B}}}\right)^5
\end{align}
This is just fast enough to avoid cosmological constraints. 
\end{itemize}
In both cases, we could then have a natural unification of the Peccei-Quinn scale and $\MS$. 

\section{Conclusions}
\label{SEC:Conclusions}

In this paper, we have further investigated Fake Split-Supersymmetry Models (FSSM) going beyond the cases introduced in~\cite{FSSM1, dudas_flavour_2013}. The main motivation is their extremely robust prediction of the correct Higgs mass in an impressive range of values of the SUSY scale $\MS$, something that can not be obtained in the original Split SUSY or High scale SUSY models.

We have shown that consideration of models where only the higgsinos are fake but not gauginos allows the retention of the main features of the FSSM with less stringent constraints. The UV completion of this new model involves a small number of additional matter fields and the hierarchy in the spectrum is ensured by an approximate R-symmetry. It is very different to -- and much more conservative than --  the original FSSM-I which in the UV is  a Dirac gaugino model with an extra flavour symmetry.

Next, we implemented both models, along with their UV completions in a code to determine the pole Higgs mass and all of the spectrum at low-energy. Once again, we stress that the Higgs mass prediction in these models is very robust. For unified masses at the GUT scale, $\tan \beta \sim 1$, all SUSY scales above $10^9$ GeV  give a $125$ GeV Higgs. If one allows values of $\tan \beta$ between $1$ and $5$, we have show that a $125$-Gev Higgs can  be ``predicted'' without constraints on the SUSY scale as can be seen in Figure~\ref{fig:mh_msusy}. We have also checked that unification was preserved  at a percent level at two-loops (see Figure~\ref{fig:unif_msusy}).

Finally we have considered the cosmology of the FSSM, extending the outline in~\cite{FSSM1}. We have distinguished three dark matter scenarios, a pure (F-)Wino with mass  $\sim2400 $ GeV in the scenario $\WDM$, and pure F-higgsino with mass in  $\sim1100$ GeV in the scenario $\HDM$, and finally a mixed (F-)Bino/F-higgsinos with close pole masses in the scenario $\BHDM$; this latter scenario exhibits rather different behaviour to equivalent coannihilation regions in other theories such as Split SUSY or the MSSM. We have found that, as in Split SUSY, if one insists on having the gluino liftime shorter than $100$s in order to preserve Big Bang Nucleosynthesis, then the SUSY scale is bounded below $5 \cdot 10^{8}$ GeV in the FSSM-I and below $2 \cdot 10^{10}$ GeV in the FSSM-II. Direct detection experiments can also constrain the FSSM for a F-higgsino LSP. Indeed F-higgsinos are a good representatives of ``inelastic'' dark matter since their splitting is suppressed by the same approximate symmetry which protect their masses. Current bounds were found to constrain the SUSY scale below  $5 \cdot 10^{6}$ GeV for the FSSM-I and below $10^{8}$ GeV for the FSSM-II. Further improvements in these experiments will translate directly into strong bounds on $\MS$ since the splitting between the F-higgsinos depends only linearly (or quadratically) on it (see eq.~(\ref{eq:approxsplit})). The cosmological constraints are summarised in Table~\ref{table:constraints}.

\section*{Acknowledgments}
We thank Pietro Slavich for useful discussions. M.~D.~G. would like to thank Luis Iba\~nez for stimulating comments and Luis Aparicio for several interesting discussions. K.~B.~ acknowledges support from the ERC advanced grant ERC Higgs@LHC. This work is supported in part by the Institut Lagrange de Paris. 

\newpage

\section*{Appendices}
\addcontentsline{toc}{section}{Appendix}

\begin{Appendix}
\label{sec:appendix}

\subsection{Implementation}
\label{APP:implementation}

The Higgs mass along with the low-energy spectrum are computed using a two-fold procedure. On one side, we compute the running between  the TeV scale and $\MS$. On the other side we compute
the running between $\MS$ and the unification scale. The consistency of the computation is insured through proper matching of the boundary conditions at $\MS$.

Running parameters between the electroweak and the SUSY scale are obtained using the code described in \cite{FSSM1} where boundary conditions are imposed both at $\MS$ to match the SUSY region predictions and at the electroweak scale to match the SM inputs. RGEs are then solved iteratively (using numerical routines from SPheno~\cite{porod_spheno_2003,porod_spheno_2012}) until we reach a solution satisfying both boundary conditions at the required precision. 

The RGEs above the SUSY scale have been obtained using the public code \SARAH~(see ref.~\cite{staub_sarah_2008,staub_automatic_2011,staub_superpotential_2010,staub_sarah_2013,staub_sarah_2014} and ref.~\cite{goodsell_two-loop_2013}).  

Our input parameters are the following
\begin{itemize}
\item The F-Higgsino $\mu$-term, $\mu_{f}$
\item The true Higgsino $\mu$-term, $\mu \sim \MS$
\item The unified F-gaugino Majorana mass $m_{fg}$ and the usual unified gaugino mass $M_{1/2}$. In the FSSM-I, only the F-gaugino mass is at the TeV scale while the gauginos are at the SUSY scale. In the FSSM-II, the gaugino mass is suppressed down to the TeV scale as seen in the previous section.
\item The SUSY scale $\MS$, which also serves as a unified mass scale for all SUSY-breaking scalar mass terms (but those for the Higgs doublet in the NUHM case)
\item The unified trilinear coupling $A_0$.
\end{itemize}
The small parameter $\varepsilon$ is defined from the (F-)gauginos mass
\begin{equation}
 \varepsilon = \begin{cases} 
                \sqrt{\frac{m_{fg}}{M_S}}  & \text{ in the FSSM-I} \\
		\sqrt{\frac{M_{1/2}}{M_S}}  & \text{ in the FSSM-II} \\
               \end{cases}
\end{equation}
so that the mass of the light gauginos-like particle is $\scr{O}(M_s \varepsilon^2) \simeq  \scr{O}(1)$ TeV.

Since the low-energy spectrum contains only F-higginos and (F-)gauginos, most of the parameter space in the UV is redundant. As a simplifying assumption, we use $\mu$ as a common scale for all unsuppressed superpotential $\mu$-like and $B_\mu$-like terms, $\mu_{f}/\varepsilon$ for all superpotential terms $\varepsilon$-suppressed and $\mu_{f}$ for the $\varepsilon^2$-suppressed terms. 

One subtlety is that even if the F-higgsinos are to leading order in $\varepsilon$ directly derived from their UV counterparts, their masses should formally be obtained by diagonalising the mass matrix for the higgsino-like particles. In order to make sure that our simplifying assumptions do not turn into fine-tuning (which happens when the determinant of the mass matrix becomes zero), we made the following choice in the FSSM-I (the FSSM-II being free from this issue): the F-higgsino $\mu$-term is $\mu_f$ and the mixing between fake and usual Higgs doublets are defined as $\frac{\mu_f}{5 \varepsilon} $. We take $B_{\mu_f} = \mu_f^2$. This choice does not modify the low-energy physics and allows us to make sure that $\mu_f$ really controls the mass of the F-higgsinos in the low-energy theory.

A similar issue arises when diagonalising the gaugino mass matrix, so in the FSSM-I the gauginos' Dirac masses are defined suppressed by a loop factor at $ \frac{1}{16 \pi^2} m_{fg}$. This choice similarly allows us to make sure that $ m_{fg}$ controls the mass of the F-gauginos in the low-energy theory.

The $B_\mu$-term for the Higgs doublets is fixed at the SUSY scale by the requirement of having a light SM-like Higgs
\begin{equation}
 B_\mu \simeq \sqrt{ (m_{H_u}^2 + \mu_{u}^2)(m_{H_d}^2 + \mu_{d}^2)} 
 \end{equation}
where $ \mu_{u}= \mu_{d}=\mu $ in the FSSM-I case and we have neglect $\epsilon$-suppressed contributions.

We take the top pole mass to be $M_t = 173.34 \pm 0.76$ GeV~\cite{atlas_and_cdf_and_cms_and_d0_collaborations_first_2014} and the strong gauge coupling to be $\alpha_3 (M_Z) = 0.1184 \pm 0.0007$~\cite{bethke_world_2012}. We use the experimental Higgs mass $M_h = 125.09 \pm 0.24$ from the combined ATLAS and CMS results~\cite{aad_combined_2015}.

The light eigenstates are predominantly composed of the original Higgs doublets and contain fake doublets only at $\scr{O} (\varepsilon)$. Hence, the mixing angle $\beta$ is given by
\begin{equation}
\label{tanbeta}
 \tan \beta  = \sqrt{\frac{m_{H_d}^2 + \mu_{u}^2}{m_{H_u}^2 + \mu_{d}^2}} \ ,
\end{equation}
and it is used  to parameterise the Higgs observables, mass and Yukawa couplings. The variation of  $ \tan \beta$ allows to reproduce the cases with $ \mu_{u} \neq \mu_{d}$ as well as non-universal Higgs masses (NUHM) set-up, where $m_{H_d}^2$ and $m_{H_u}^2$  have different values at $M_{GUT}$.

Supersymmetry predicts the SM-like tree-level Higgs quartic coupling at $\MS$ via equation (\ref{lambdaMS}):
\beq
\lambda (\MS)~=~
\frac14\left(g^2+g^{\prime\,2}\right)\,\cos^22\beta ~+~ \Delta^{(\ell)} \lambda ~+~ \Delta^{(\ov{MS})} \lambda~+~{\cal
  O}(\varepsilon^2).
\eeq
 The corrections $\mathcal{O}(\varepsilon)$ are always negligible in this work, however the loop contributions can play a role. At one loop, we have the leading stop contribution given by
\begin{align}
 \Delta^{(1)} \lambda \supset \frac{3 y_t^4}{16\pi^2} \bigg[\log \frac{m_{Q_3}^2 m_{U_3}^2 }{\MS^4} + \mathcal{O}(\tilde{X}_t)\bigg]
\end{align}
where $y_t$ is the top Yukawa coupling, $\tilde{X}_t \equiv \frac{|A_t - \ov{\mu} \cot \beta|^2}{m_{Q_3} m_{U_3}} $ and the dependence on this can be found e.g. in \cite{bagnaschi_higgs_2014}. Since the stop contribution is the most important, we make the standard convenient choice of using it to define $\MS \equiv \sqrt{m_{Q_3} m_{U_3}}$. In the FSSM-II , $A_t$ and $\mu$ term are suppressed by the R-symmetry, so we can safely take $\tilde{X}_t \simeq 0$. In the FSSM-I, however,  both are in general quite large; we have estimated the shift of the Higgs mass to be at most  $4.5$ GeV when  when $\MS \sim 100$ TeV and at most  to $1$ GeV when $\MS \sim 10^8$ GeV. In most of our plots, $A_t$ and $\mu$ are chosen to be equal to $\MS$ at the GUT scale so the shift is further reduced to circa $2$ GeV even for  $\MS \sim 100$ TeV.

Other threshold corrections include terms from decoupling the heavy MSSM particles and changes of the renormalisation schemes from $\overline{D \! R}$ to  $\overline{M \! S}$. For the case of Split SUSY, the expressions are given in~\cite{bagnaschi_higgs_2014}. We have found the effects in our models to lead to a sub-GeV contribution to the Higgs mass so they have been neglected; however it would be interesting to be able to compute these contributions for our model to completely assess their effect.

\subsection{Dirac dark matter from Fake Split Extended Supersymmetry and the $3.55$ keV line}
\label{APP:FSES}

Over the past year there was much attention given to the possibility that a $3.5$~keV line observed in combining 73 galaxy clusters \cite{Bulbul:2014sua} and in the Perseus cluster (and Andromeda galaxy) \cite{Boyarsky:2014jta} may originate from dark matter decay. 
It was initially interpreted in terms of sterile neutrino decay, as the mass and signal strength sit in the allowed/predicted window for such particles to constitute dark matter. However, since the initial excitement there have been challenges to the decaying dark matter interpretation \cite{Urban:2014yda,Carlson:2014lla,Iakubovskyi:2014yxa}, including from the non-observation of the line in stacked dwarf spheroidal galaxies \cite{Malyshev:2014xqa}  and other stacked galaxies \cite{Anderson:2014tza} despite its observation in the Milky Way \cite{Boyarsky:2014ska}. Perhaps the most plausible explanations that avoid these issues are excited dark matter \cite{Finkbeiner:2007kk,Pospelov:2007xh,Finkbeiner:2014sja,Frandsen:2014lfa,Cline:2014vsa} and an dark matter decaying to an axion-like particle in the magnetic field of a cluster \cite{Cicoli:2014bfa,Conlon:2014xsa,Kraljic:2014yta,Conlon:2014wna,Alvarez:2014gua}. On the other hand, the decaying dark matter explanation is not yet completely excluded, and so in this appendix we shall describe how a class of models related to the FSSM provides an explanation for the line.

To produce a line from a fermion $\Psi_2$ that decays to a photon and another fermion $\Psi_1$ (with two-body decays preferred to give a sharp line) we require either a large difference in the masses, $m_2 \gg m_1$, as for a sterile neutrino, or the difference to be equal to the photon energy, $m_2 - m_1= 3.55$ keV, as in e.g. \cite{Falkowski:2014sma}. Clearly in the FSSM we do not expect an extremely light neutralino, and so the latter explanation is preferred. Since the fermion is neutral, we shall take them to be Majorana and their coupling with photons should be of dipole type:
\begin{align}
\mathcal{L} \supset& ~\ov{\Psi}_2 \gamma^{\mu \nu} (C_{12} P_L + C^*_{12} P_R) \Psi_1 F_{\mu \nu} 
\end{align}
which mediates $\Psi_2$ decay to $\Psi_1$ with a rate 
\begin{align}
\Gamma =& ~|C_{12}|^2 \frac{(m_2^2 - m_1^2)^3}{2\pi m_2^3}.
\end{align}
The mass splitting should be equal to $3.55$~keV; to explain this near-degeneracy we expect to evoke an approximate symmetry where an initially Dirac fermion is broken to two Majorana eigenstates. 
The required value of $C_{12}$ to explain the line is given by
\begin{align}
C_{12} \simeq& ~5 \times 10^{-15}\ \mathrm{GeV}^{-1}\ \left( \frac{m_2}{100\ \mathrm{GeV}} \right)^{1/2}.
\end{align}
Let us denote the width of a $7$ keV particle decaying to a photon and a near-massless particle which would match the observed line as 
\begin{align}
\Gamma_\nu \simeq& ~1.1 \times 10^{-52}\ \mathrm{GeV}.
\end{align}
Then the width required by our particle, defined as $\Gamma_2 $, is
\begin{align}
\Gamma_2 =& ~\frac{2 m_2}{7\ \mathrm{keV}} \times \Gamma_\nu \nn\\
\simeq& ~0.3 \times 10^{-43}\ \mathrm{GeV}\, \times \left(\frac{ m_2}{\mathrm{TeV}} \right) .
\end{align}
where the factor of $2$ is due to there being two dark matter particles assumed to be of near-equal density, but only one radiates. We could imagine that this particle does not make up all of the dark matter in the universe, but only some fraction, and instead has a larger width still. However, we rapidly come across a barrier to this: the decay rate should not be so fast that its lifetime is less than the age of the universe,
\begin{align}
\tau_{\mathrm{Universe}}^{-1} =& 1.5 \times 10^{-42}\, \mathrm{GeV}.
\end{align}
Hence a dark matter particle at a TeV is already starting to approach this limit and we should consider that it makes up a substantial fraction of the dark matter. This also places an upper limit on the mass of the dark matter particle. 

\subsubsection{Fake Split Extended Supersymmetry}

If $C_{12}$ is generated by loops of heavy particles coupling to the Majorana fermions with strength $\lambda$, then in the limit of large masses $M$ we find 
\begin{align}
C_{12} \simeq& \frac{\lambda^2 e}{32\pi^2 M} \nn\\
\frac{M}{\lambda^2} \sim& 10^{11}\ \mathrm{GeV}. 
\end{align}
This hints at new physics at an intermediate scale (or rather weakly coupled $\lambda \sim 10^{-4}$ at $M \sim $ TeV) which could be naturally related to the  (Fake) Split Supersymmetry scale. However, the FSSM does not have a natural pseudo-Dirac femion that could explain the line, since a pseudo-Dirac (fake) higgsino with such a small mass-splitting between its neutral components is thoroughly ruled out as a dark matter candidate by direct detection constraints, and in addition would decay preferentially to neutrinos via the Z much faster than the age of the universe:
\begin{align}
\Gamma(h_2 \rightarrow h_1 \ov{\nu}\nu) \simeq& 3 \times \frac{\alpha^2 m_2^5 (1-2 s_W^2)^2 (\delta m)^5}{40 \pi c_W^2 s_W^4 M_W^2 M_Z^2 } \simeq 10^{-30}\ \mathrm{GeV} \left( \frac{m_2}{100\ \mathrm{GeV}}\right)^5.
\label{EQ:higgsinorate}\end{align}
 Instead we shall introduce here a new theory at the electroweak/TeV scale with \emph{Dirac} gauginos and fake higgsinos.

Our model is a slight modification of \emph{Split Extended Supersymmetry} \cite{Antoniadis:2005em, Antoniadis:2006eb} (see also \cite{Carena:2004ha,Unwin:2012fj,Fox:2014moa} for related work) where we add additional states to ensure unification of gauge couplings -- and also replace the higgsinos with F-higgsinos. We know that if we start with the CMDGSSM matter content \cite{benakli_constrained_2014} and make the scalars heavy, then we will preserve unification: to be more explicit, let us compute the shift in the beta functions. For regular Split-SUSY compared to the MSSM we have:
\begin{align}
\Delta b_3^{\mathrm{Split\ SUSY}} =& \underbrace{\frac{1}{3}}_{\mathrm{scalars\ only}} \times \bigg(  \underbrace{3\times \frac{1}{2} \times 2}_{Q} + \underbrace{3\times \frac{1}{2} \times 2}_{U+D} \bigg) = 2\nn\\
\Delta b_2^{\mathrm{Split\ SUSY}} =& \frac{1}{3} \times \bigg(  \underbrace{3\times 3 \times \frac{1}{2} }_{Q} + \underbrace{3\times \frac{1}{2} }_{L} +  \underbrace{2\times \frac{1}{2} }_{H_{u,d}}-  \underbrace{\frac{1}{2} }_{H}\bigg) = 2 + \frac{1}{6}\nn\\
\Delta b_1^{\mathrm{Split\ SUSY}} =& \frac{1}{3} \times \frac{3}{5} \times \bigg(  \underbrace{3\times 3 \times 2 \times\frac{1}{36} }_{Q} +  \underbrace{3\times 3 \times \frac{4}{9} }_{U} \nn\\
& +  \underbrace{3\times 3 \times \frac{1}{9} }_{D} + \underbrace{3\times 2 \times \frac{1}{4} }_{L} + \underbrace{3 \times 1 }_{E} +  \underbrace{2\times 2 \times  \frac{1}{4} }_{H_{u,d}}-  \underbrace{2\times \frac{1}{4} }_{H}\bigg) = 2 + \frac{1}{10} 
\end{align}

For our new scenario -- Fake Split Extended Supersymmetry -- the shift relative to Split SUSY is
\begin{align}
\Delta^\prime b_3 =& \frac{1}{3} \times \bigg( \underbrace{3}_{O} \bigg) = 1 \nn\\
\Delta^\prime b_2 =& \frac{1}{3} \times \bigg( \underbrace{2}_{T}  + \underbrace{2\times \frac{1}{2} }_{R_{u,d}}\bigg) = 1\nn\\
\Delta^\prime b_1 =&  \frac{1}{3} \times \frac{3}{5} \times \bigg(  \underbrace{2\times 2 \times  \frac{1}{4} }_{R_{u,d}} + \underbrace{2 \times 2 \times 1 }_{\hE,\hEt} \bigg) = 1
\end{align}
and hence unification is just as good as split SUSY. 

In this scenario, however, in order to preserve Dirac gauginos at low energy \emph{and} keep some other states light, we must have both an approximate R-symmetry and a $U(1)_F$ symmetry. The breaking of $U(1)_R$ should be much smaller, so that the Majorana masses induced are of order $3.55$ keV to explain the line. Then the field content and charges at $\MS$ is the set of MSSM matter fields plus 
\begin{center}
\begin{tabular}{|c|c|c|c|}\hline
Superfield & $(SU(3), SU(2), U(1)_Y) $ & $R$ & $F$ \\\hline
$\mathbf{O}$ & \reps[8,1,0] & $0$ & $2$ \\ \hline
$\mathbf{T}$ & \reps[1,3,0] & $0$ & $2$ \\ \hline
$\mathbf{S}$ & \reps[1,1,0] & $0$ & $2$ \\ \hline
$\mathbf{H_u}, \mathbf{H_d}$ & \reps[1,2,\pm \frac{1}{2}] & $0$ & $0$ \\ \hline
$\mathbf{R_u}, \mathbf{R_d}$ & \reps[1,2,\mp \frac{1}{2}] & $2$ & $0$ \\ \hline
$\mathbf{H_u^\prime},\mathbf{H_d^\prime}$ & \reps[1,2,\pm \frac{1}{2}] & $0$ & $1$ \\ \hline
$\mathbf{R_u^\prime}, \mathbf{R_d^\prime}$ & \reps[1,2,\mp \frac{1}{2}] & $2$ & $1$ \\ \hline
$\mathbf{\hE}_i,\mathbf{\hEt}_j$ & \reps[1,1,\pm 1] & $1$ & $1$\\ \hline
\end{tabular}\end{center}
As in Split Supersymmetry, all of the scalars obtain masses at $\MS$, and all of the fermions have masses suppressed by the breaking of $U(1)_F$ to $\mathcal{O}(\mathrm{TeV})$ \emph{except} for those made massive by the superpotential
\begin{align}
\mathbf{W^{\mathrm{Fake\ Extended}}} \supset \mu_u \mathbf{H}_u \mathbf{R}_d + \mu_d \mathbf{R}_u \mathbf{H}_d + \mathcal{O}(\varepsilon).
\end{align}
This scenario will then give different predictions for the Higgs mass compared to the FSSM. We have implemented the RGEs (using \SARAH) in an adapted version of our code for this model and undertaken a very preliminary scan, shown in figure \ref{DiracSplitHiggs}. Interestingly, this model retains the prediction of consistency with the observed Higgs mass for \emph{any} value of $\MS$ but with larger $\tan \beta$ (defined via the mixing between the Heavy Higgses $H_u, H_d$ at $\MS$). We leave however a more thorough investigation for future work.

\begin{figure}\begin{center}
\includegraphics[width=0.6\textwidth]{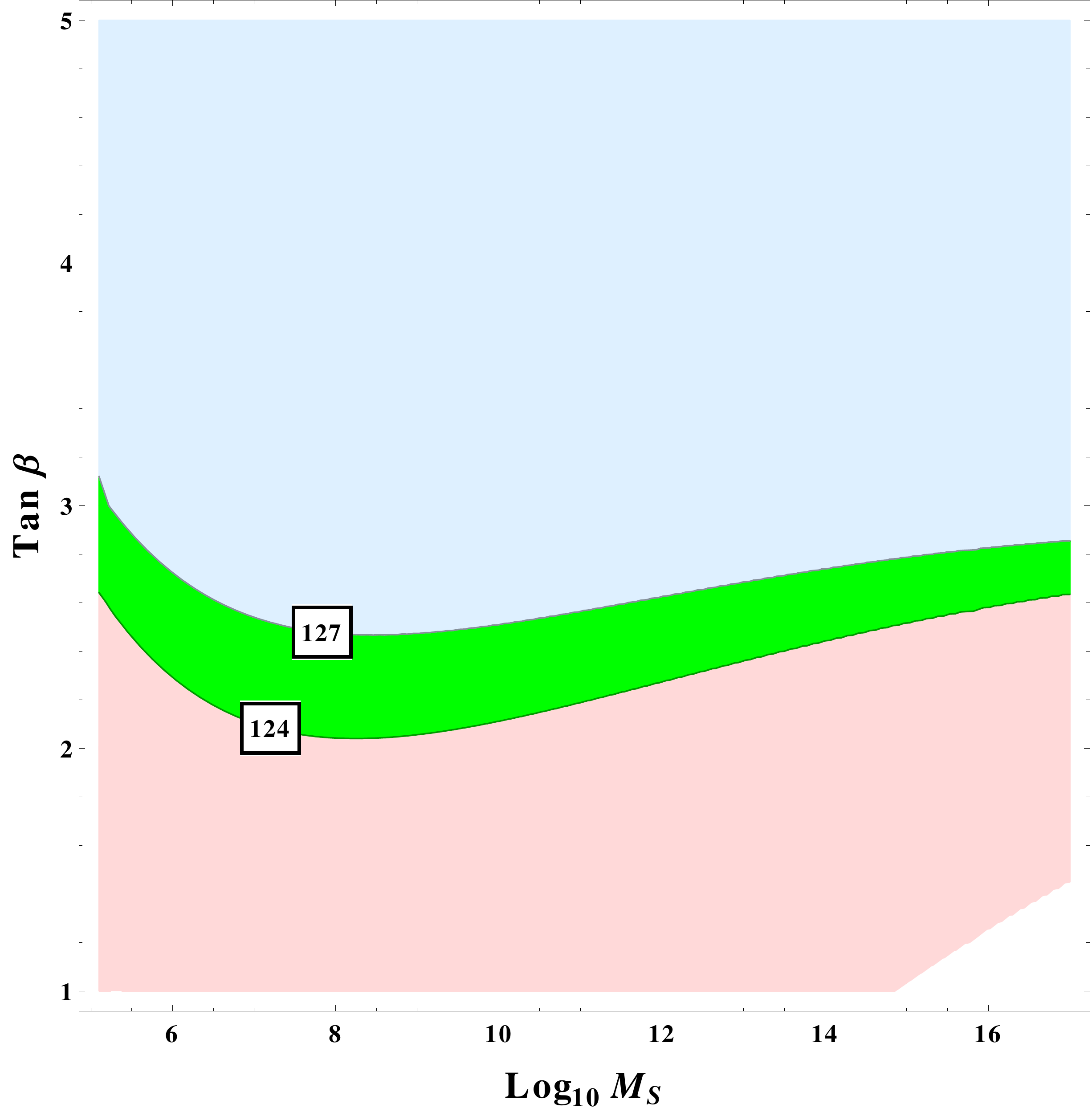}
\caption{Contours of Higgs mass against SUSY-breaking scale and $\tan \beta$ in the Fake Split Extended SUSY scenario.}
\label{DiracSplitHiggs}
\end{center}\end{figure}

\subsubsection{X-ray line candidates in Fake Split Extended Supersymmetry}

This model possesses four neutral pseudo-Dirac fermions: the Bino, Wino and two F-higgsinos. As described above the F-higgsinos are excluded as a description for the line. For the other candidates, the issue is whether the dipole operator will be \emph{small} enough; we require that low-energy processes will not generate the operator which would then only by suppressed by $M_W$ rather than $\MS$. For the Bino, we find that the dipole operator is generated at one loop from interactions with both heavy states (squarks, sleptons etc) and also through mixing with the higgsinos, but the magnitude is 
\begin{align}
C_{12} \sim \frac{g_Y^3}{16\pi^2 \MS} \sim \frac{g_Y^3 \varepsilon^2}{16\pi^2 \mathrm{TeV}}.
\label{EQ:binodipole}\end{align}  
This is then consistent with the observed line if $\MS \sim 10^{12}$ GeV. To populate the correct relic density we require the Bino and higgsino to have similar masses as in the FSSM, but just as in that case the mixing between the two states is still small (only enhanced by one or two orders of magnitude if the mass difference is of order the freezeout temperature) and therefore does not spoil the prediction for $C_{12}$. We refer the reader to the discussion in section \ref{sec:RelicDensity}.

Alternatively, we could have Wino dark matter without requiring similar masses for the Bino and higgsino. For a Dirac Wino, we have a neutral Dirac fermion and \emph{two} Dirac charginos. Naively we expect that loops involving charginos would generate $C_{12} \sim \frac{g_2^2 e}{16\pi^2 M_W}$ (clearly higgsino loops, since they are suppressed by mixing, generate an operator of magnitude given by equation (\ref{EQ:binodipole})). However, neglecting the mixing between Winos and higgsinos (since this is $\varepsilon$-suppressed) if only Dirac masses are present -- in the absence of R-symmetry breaking -- the charginos are degenerate and with opposite signs. The contributions to the dipole operator then cancel out. This persists to all orders, because it leads to a residual symmetry upon exchanging the Wino (Weyl) states with their corresponding (Weyl) fermion of the same charge, under which the dipole operator is odd. Hence the dipole operator must be proportional to the breaking of this symmetry, i.e. 
\begin{align}
C_{12}^{\rm wino} \sim  \frac{g_2^2 e }{16\pi^2 M_W} \times \frac{\mathrm{keV}}{M_W} \sim 10^{-13}\ \mathrm{GeV}.
\end{align}
This is rather close to the required value; if we had the Wino mass in the denominator then we would find $10^{-15}$ GeV, a remarkable coincidence.

\end{Appendix}

%
\newpage

\bibliographystyle{unsrt}
\bibliography{FSSM}

\end{document}